\documentclass[twocolumn]{aastex62}
\usepackage{longtable}
\usepackage{verbatim}
\usepackage{chngcntr}

\graphicspath{{./}{figures/}}

\newcommand{\civ}{C~{\small IV}}
\newcommand{\heii}{He~{\small II}}
\newcommand{\ciii}{C~{\small III}]}
\newcommand{\oiii}{[O~{\small III}]}
\newcommand{\nv}{N~{\small V}}

\newcommand{\civl}{\civ~$\lambda$1549}
\newcommand{\heiil}{\heii~$\lambda$1640}
\newcommand{\ciiil}{\ciii~$\lambda$1909}
\newcommand{\nvl}{\nv~$\lambda$1240}
\newcommand{\oiiil}{\oiii~$\lambda\lambda$4959,5007}

\received{}
\revised{}
\accepted{}
\submitjournal{ApJ}

\shorttitle{Metal enrichment of the circumgalactic medium of Quasars}
\shortauthors{Guo et al.}

\begin{document}

\title{Metal enrichment in the circumgalactic medium and Ly$\alpha$ haloes around  quasars at $z\sim3$}

\author{Yucheng Guo}
\affiliation{Kavli Institute for Astronomy and Astrophysics, Peking University, Beijing 100871, China}
\affiliation{Department of Astronomy, School of Physics, Peking University, Beijing 100871, China}
\affiliation{Cavendish Laboratory, University of Cambridge, 19 J. J. Thomson Ave., Cambridge CB3 0HE, UK}
\affiliation{Kavli Institute for Cosmology, University of Cambridge, Madingley Road, Cambridge CB3 0HA, UK}

\author{Roberto Maiolino}
\affiliation{Cavendish Laboratory, University of Cambridge, 19 J. J. Thomson Ave., Cambridge CB3 0HE, UK}
\affiliation{Kavli Institute for Cosmology, University of Cambridge, Madingley Road, Cambridge CB3 0HA, UK}

\author{Linhua Jiang}
\affiliation{Kavli Institute for Astronomy and Astrophysics, Peking University, Beijing 100871, China}

\author{Kenta Matsuoka}
\affiliation{
Research Center for Space and Cosmic Evolution, Ehime University, 2-5 Bunkyo-cho, Matsuyama, Ehime 790-8577, Japan
}

\author{Tohru Nagao}
\affiliation{Research Center for Space and Cosmic Evolution, Ehime University, 2-5 Bunkyo-cho, Matsuyama 790-8577, Japan}

\author{Oli Luiz Dors}
\affiliation{Universidade do Vale do Para{\'i}ba. Av. Shishima Hifumi, 2911, CEP: 12244-000, S{\~ a}o Jos{\'e} dos Campos, SP, Brazil}

\author{Michele Ginolfi}
\affiliation{Observatoire de Gen\`eve, Universit\`e de Gen\`eve, 51 Ch. des Maillettes, 1290 Versoix, Switzerland}

\author{Nick Henden}
\affiliation{Institute of Astronomy, Madingley Rd, Cambridge CB3 0HA, UK}
\affiliation{Kavli Institute for Cosmology, University of Cambridge, Madingley Road, Cambridge CB3 0HA, UK}

\author{Jake Bennett}
\affiliation{Institute of Astronomy, Madingley Rd, Cambridge CB3 0HA, UK}
\affiliation{Kavli Institute for Cosmology, University of Cambridge, Madingley Road, Cambridge CB3 0HA, UK}

\author{Debora Sijacki}
\affiliation{Institute of Astronomy, Madingley Rd, Cambridge CB3 0HA, UK}
\affiliation{Kavli Institute for Cosmology, University of Cambridge, Madingley Road, Cambridge CB3 0HA, UK}

\author{Ewald Puchwein}
\affiliation{Leibniz-Institut f\"ur Astrophysik Potsdam,
An der Sternwarte 16, 14482 Potsdam, Germany}

\begin{abstract}
Deep observations have detected extended Ly$\alpha$ emission nebulae surrounding tens of quasars at redshift 2 to 6. However, the metallicity of such extended haloes is still poorly understood. We perform a detailed analysis on a large sample of 80 quasars at $z\sim3$ based on MUSE-VLT data. We find clear evidence of extended emission of the UV nebular lines such as \civl\ or \heiil\ for about 20$\%$ of the sample, while \ciiil\ is only marginally detected in a few objects. By stacking the cubes we detect emission of \civ, \heii\ and \ciii\ out to a radius of about 45 kpc. \civ\ and \heii\ show a radial decline much steeper than Ly$\alpha$, while \ciii\ shows a shallower profile similar to Ly$\alpha$ in the inner 45 kpc. We infer that the average metallicity of the circumgalactic gas within the central 30--50~kpc is $\sim$0.5 solar, or even higher. However, we also find evidence of a component of the Ly$\alpha$ haloes, which has much weaker metal emission lines relative to Ly$\alpha$. We suggest that the high metallicity of the circumgalactic medium within the central 30--50 kpc is associated with chemical pre-enrichment by past quasar-driven outflows and that there is a more extended component of the CGM that has much lower metallicity and likely associated with near-pristine gas accreted from the intergalactic medium. We show that our observational results are in good agreement with the expectations of the FABLE zoom-in cosmological simulations.
\end{abstract}


\section{Introduction}

Within the circumgalactic medium (CGM), gas and metals are ejected from galaxies by feedback processes, or stripped from infalling satellites. The CGM also hosts the reservoir of metal poor gas accreted from the intergalactic medium (IGM) that can eventually accrete on galaxies to fuel star formation. The matter and energy exchange between the IGM, CGM and galaxies is critical in understanding galaxy star formation, enrichment history and morphological type. Investigating the CGM and understanding its connection to the star formation activity and interstellar medium (ISM) in galaxies are important to understand galaxy evolution.

A common method to study the CGM is the analysis of absorption signatures against bright background sources \cite[e.g.,][]{rauch11, matejek12, turner14, lehner15, rubin15, bowen16, martin19}. This method is sensitive to low column densities, and its detection limit does not depend on redshift or on host galaxy luminosity \citep{tumlinson17}. This method only provides one dimensional (1D) information, and it is limited by the sparseness of background sources. However, it still provides statistical constraints on the CGM if the samples are large enough.

Another method to study the CGM is to directly image a galaxy and its surrounding nebular emission. Directly mapping the CGM is challenging, because its emission is usually weak. Locally several studies have exploited HI 21cm radio emission or soft X-ray emission \cite[e.g.,][]{humphrey11, putman12, anderson16}. At high redshift studies have mostly focused on ultraviolet (UV)/optical wavelengths, particularly using the Ly$\alpha$ line. By using the narrow band (NB) technique, tens of individual galaxies and active galactic nucleus (AGNs) have been observed \cite[e.g.,][]{matsuda11,cantalupo14}, and diffuse Ly$\alpha$ blobs have been detected on spatial scales of tens or even hundreds of kpc. With integral field spectrograph facilities such as the Multi Unit Spectroscopic Explorer \cite[MUSE;][]{bacon10} of Very Large Telescope (VLT), tens of Ly$\alpha$ nebulae have been detected around quasars and galaxies \citep[e.g. ][]{borisova16,wisotzki16,farina17,farina19,lseclercq17,fabrizio18,cai18,ginolfi18,fabrizio19,drake19}. These studies achieved a high detection rate of Ly$\alpha$ nebulae, and provided a robust sample for studying the properties of the CGM.
Ly$\alpha$ emission is very bright, and it is physically extended due to its resonant nature. The extended Ly$\alpha$ emission could be used to detect the properties of the CGM or even the cosmic web \cite[e.g.][]{fabrizio19b,lusso19,witstok19}.
At lower redshift, similar observations have been undertaken to search for extended emission of different transitions such as H$\alpha$ or \oiiil\, \citep{schirmer13,yuma17,yuma19}.

Metals provide an important tracer to probe the interconnection between the IGM, CGM and galaxies. Metals are produced inside galaxies, but the bulk of them have been distributed outside galaxies across the cosmic epochs. \citet{peeples14} performed a census of metals in and around star-forming galaxies in the local Universe. They reported a surprisingly small fraction (20\%$-$25\%) of metals remaining in stars, interstellar gas, and interstellar dust. Recent overviews of the metal budget in and around galaxies are given in \cite{tumlinson17} and \citet{maiolino19}. Metals are transferred to the CGM via processes such as outflows and stripping \cite[e.g.][]{tremonti04,tripp11,tumlinson17}. The metal enrichment of the CGM provides information about the origin of the CGM gas. Metal poor gas in the CGM possibly suggests pristine IGM accretion \citep{chen19}, while metal-rich components of the CGM are likely associated with large-scale galactic outflows \citep{muratov17,chisholm18}.

Numerous cosmological simulations can predict the metallicity distribution and evolution in the CGM
\cite[e.g.][]{crain13,nelson18,vangioni18}. However, these predictions are subject to various
uncertainties, such as the supernova and AGN feedback efficiency, the metal loading and advection in the outflows, and mixing with the gas inflowing from the IGM. Therefore, observational constraints of the CGM metallicity are crucial to understand the physics of galaxy formation.

After Ly$\alpha$, the brightest UV emission lines from gas photoionized by quasars are \civl, \heiil\ and \ciiil\ and, sometimes \nvl\ (hereafter \civ, \heii, \ciii, \nv, respectively),
either in their host galaxy or in the CGM. Their line ratios are often used to constrain metallicity, ionization parameter, and gas density \citep{nagao06,matsuoka09,dors14,dors18,dors19,gutkin16,maseda17,matsuoka18,mignoli19}.

While previous studies have focused on the Ly$\alpha$ distribution and kinematics, in this paper we investigate the nature of the CGM in a large sample of quasars by using the strength of UV lines, e.g., \civ, \heii\ and \ciii \ (which have been little explored in previous studies), with the ultimate goal of constraining the metallicity of the CGM.
These lines are obviously much more difficult to detect and map due to their intrinsically fainter emission relative to Ly$\alpha$. Only a few  cases of extended \civ\ or \heii\ emission are found in these systems \cite[e.g.][]{borisova16,marino19}. At the redshift range of $z = 3-4$ these lines are shifted into the MUSE wavelength range. Our approach follows similar analysis methods undertaken for Ly$\alpha$ nebulae around quasars \cite[e.g.][]{borisova16,farina17,farina19,ginolfi18,fabrizio19,drake19}. We also have to rely on the use of data stacking due to the low surface brightness of these UV lines.

The structure of our paper is as follows. In Section~\ref{sec:data_analysis}, we describe our data and data reduction procedure. In Section~\ref{sec:results} we present our results. We discuss the implications of our findings in Section~\ref{sec:discussion}. Section~\ref{sec:conclusions} summarizes our results. Throughout this paper, we adopt the standard $\mathrm{\Lambda CDM}$ cosmology with $H_0\mathrm{=70\, km\, s^{-1}\, Mpc^{-1}}$, $\mathrm{\Omega _m=0.3}$ and $\mathrm{\Omega _\lambda =0.7}$. All distances are proper, unless specially noted.

\section{Data Description and Analysis}
\label{sec:data_analysis}

In this section, we analyze a large and deep MUSE sample of 80 quasars at $z\sim3$ taken from the archive. The sample combines two large samples from \citet{borisova16} and \citet{fabrizio19}.
Both papers aim to detect the Ly$\alpha$ nebulae around quasars. In our work, we focus on the other UV emission lines within the MUSE wavelength range, such as \civ, \heii, and \ciii. These emission lines are generally too weak to get individual detections in most targets, but with this large sample we can at least obtain a stacking result.
The \citet{borisova16} sample includes 17 radio-quiet quasars and 2 radio-loud quasars. The \citet{fabrizio19} sample consists of 61 quasars, including 15 radio-loud quasars. The total exposure time is $\sim$97 hrs, about 1.2 hrs for each object on average. The quasars in the sample are listed in Table~\ref{tab:sample}, along with some of their physical properties.

We downloaded all the processed and calibrated datacubes of the sample from the ESO Archive.
Residual background emission was removed from each wavelength slice of the cube by determining the average residual sky emission in a meshed grid, avoiding the central 5$''$ around the quasar.
The inferred background emission was interpolated spatially, also beneath the location of the quasar. We carefully checked that this procedure did not subtract the diffuse nebular emissions. In this procedure, we intended to eliminate the uneven background, which is insignificant in the original datacubes, but becomes relevant after subtraction of the quasar light, as discussed in Section~\ref{sec:psf_sub}.
We also masked a few very strong sky lines, as the photon noise introduced by these very bright background feature prevents the detection of any weak astronomical signal.
The masking of these sky lines has little effect on the final stacking of the cubes (which is the primary methodology adopted in our work), thanks to the shift in wavelength required to realign the spectra to their rest frame.

We note that the ESO pipeline is claimed to be sub-optimal with respect to other non-public pipelines developed by other groups \citep[e.g. those used in ][]{borisova16,fabrizio19b}, however we have verified that the final rms (especially for the stacks), after the refined background subtraction, is similar to the one obtained in those papers.

\subsection{PSF and Continuum Subtraction}
\label{sec:psf_sub}
As the emission from the central AGN is much brighter than the diffuse emission of the CGM, an important step is to remove the unresolved AGN emission from the spatially resolved emission of the CGM. To do this, we took advantage of the integral field spectroscopic information delivered by MUSE. We used the wavelength regions where we were sure that no extended line emissions should be present. To estimate and remove the quasar PSF, we adopted a purely empirical method similar to that introduced in \citet{borisova16}. For each wavelength layer, we produced a pseudo-NB image with a spectral width of $\sim$187\AA. The empirical PSF is rescaled to match the flux within the central 5$\times$5 pixels ($1''\times1''$). To avoid the contamination of cosmic rays or other artifacts, an averaged-sigma-clip algorithm was applied to calculate the scaling factor between the flux of the central quasar in each layer and the PSF images. We then cut the central circular region of the PSF image within a radius of five times the seeing, and subtracted it from the original datacube. This method provides good results, but the central $1''\times1''$ region used for PSF rescaling is not usable as the flux in this region must be zero by construction. We therefore masked the central $1''\times1''$ region in our study as we are interested in emission on larger scales. Note that the PSF at the wavelength of the nebular lines was constructed by interpolation of the PSF inferred from the nearby continuum channels.

Next, we adopted a median-filtering approach to remove continuum sources in the datacube, as described in \citet{borisova16}. This approach provides a fast and efficient way to remove continuum sources in the search for extended line emission.
We note however that the number and surface brightness of the detected haloes do not change significantly with and without median filtering.
We also masked several bright neighboring objects to avoid possible contamination.

The subtraction of PSF and continuum emission was done for each exposure. We stacked different exposures of the same object, by spatially re-aligning the cubes, weighted by their exposure time.

In order to improve the signal to noise (S/N) and to obtain the average radial profile we also adopted two different stacking methods that will be described in Section~\ref{sec:results}. As already mentioned for the stacking, in order to further improve the sensitivity, we have masked the wavelengths affected by strong OH lines, prior to the shifting to the wavelength rest frame and prior to averaging the cubes.

\subsection{Line Emission Extraction} \label{subsec:data_3dextract}
The final step is to extract and identify the extended line emission from the PSF and continuum subtracted datacubes.
The extraction was performed based on a S/N threshold on each individual spectral slice of the cube. Voxels with S/N lower than 1 were masked. Note that this is a S/N threshold much more conservative than that used in other works. For instance, \cite{borisova16} adopt a S/N threshold of 2 on spaxels spatially smoothed with a Gaussian filter with a $\sigma = 0.5''$, which is equivalent to a S/N threshold of 0.34 on individual unsmoothed spaxels. The S/N threshold results into a three dimensional (3D) mask for each datacube and, finally to obtain a ``S/N clipped" datacube. Then we applied pseudo-NB filters to the wavelength ranges of \civ, \heii, and \ciii\ by summing up the flux along the wavelength axis for pixels selected by the 3D masks (i.e. only those pixels meeting the S/N threshold requirement in each slice). The width of each pseudo-NB filter is about $\mathrm{30\, \AA}$ in rest-frame around the line centre. Finally we obtain the pseudo-NB image of each emission line. We note that the adopted width of the pseudo-NB filter
is always significantly larger than the width of the nebular line emission in the halo (as it will be shown in Section~\ref{subsec:results_kinematicss}). Moreover, we emphasize that the pseudo-NB filter is applied to the ‘S/N clipped datacube’, hence even if larger than the line it does not introduce much extra-noise from layers or spaxels not containing signal, as these have been filtered out.

We verified that the UV lines detected (or even marginally detected at 2$\sigma$) are not residuals of the quasar light PSF subtraction by verifying that they are much narrower than the quasar emission lines, that they are not spatially
radially symmetric, and that they are well above the signal measured in the neighbor continuum emission profile. These
tests will be clarified further especially in the case of the stacked emission.

An accurate systemic redshift is important to detect UV line emission and kinematics and also for accurately shifting the spectra to the same rest-frame for stacking multiple spectra and cubes from different quasars. For quasars at $z>2$, an accurate measure of systemic redshift is hard to achieve, and generally relies on the observations in near infrared or molecular tracers \cite[e.g.,][]{mcintosh99,venemans17}. We could get the corrected redshift estimated from the broad \civ\ line, but the intrinsic uncertainty is about $\mathrm{\sim 400 \, km\, s^{-1}}$ \citep{shen16}, which is too large for the requirements of diffuse line detection. Therefore,
in this paper, we used the redshift estimated from the diffuse Ly$\alpha$ nebulae, i.e. the narrow and extended component of Ly$\alpha$.
For the objects that have clear individual detection of diffuse \civ, \heii, or \ciii\ nebulae, we found that the redshifts of diffuse Ly$\alpha$ nebulae (calculated by the spectrum extracted in $3\arcsec$) is consistent with those of \civ, \heii, or \ciii\ within the error range. This is not unexpected, given that any gas producing any of the CIV, HeII or CIII] must also produce Ly$\alpha$ by recombination. The consistency of the velocities of these emission lines with Ly$\alpha$ narrow will be further discussed in Section~\ref{subsec:results_kinematicss}. Therefore, summarizing, we use the redshifts of diffuse Ly$\alpha$ as the systemic redshifts of the quasars.

Note that also for the extraction of the line maps the wavelengths of strong OH lines were masked to improve the sensitivity.

\section{Results}
\label{sec:results}
\subsection{Extended UV Emission Line Haloes} \label{subsec:results_halos}
\begin{figure}[ht!]
\includegraphics[width=0.5\textwidth]{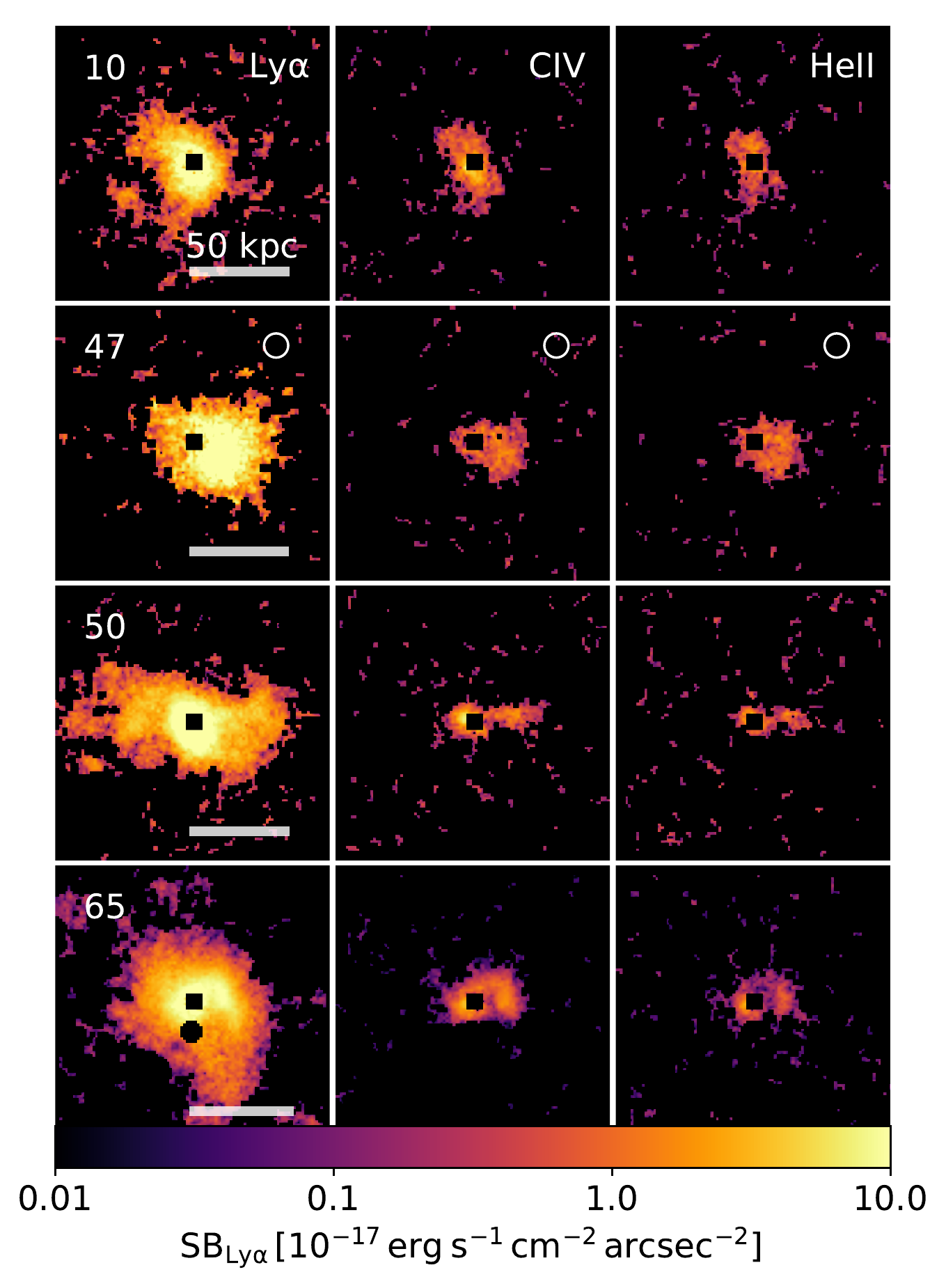}
\caption{Maps of the Ly$\alpha$, \civ\ and \ciii\ emission for four quasars that show extended emission of both \civ\ and \heii\ (No. 10, 47, 50, 65 in Table~\ref{tab:sample}). In each panel, the original position of each quasar is masked by a small ($1''\times1''$) black square. Each thumbnail has a size of $20''\times20''$, corresponding to about 150 kpc at the median redshift of the sample. These images are produced by collapsing S/N-clipped datacubes with 3D masks along the wavelength direction. The black circle in the lower left panel denotes a masked bright star.
\label{fig:linemap_objs}}
\end{figure}

Following the reduction steps in Section~\ref{sec:data_analysis}, we detect Ly$\alpha$, \civ, \heii, \ciii\ nebulae extended on circumgalactic scales around a subset of individual quasars in our sample. Line maps of four objects for which we can detect and resolve both \civ\ and \heii\ are shown as examples in Figure~\ref{fig:linemap_objs}. The scale of the thumbnails is $20\arcsec  \times 20\arcsec$ (corresponding to 150 $\times$ 150 kpc at the median redshift of the sample). For convenience, the number on the top left corner of each image indicates the ID of each targets listed in Table~\ref{tab:sample}. The atlas of emission line maps of all detected nebulae are shown in Figures \ref{fig:lya_halos}, \ref{fig:civ_halos}, \ref{fig:heii_halos} and \ref{fig:ciii_halos} in Appendix~\ref{appen:pseudoNB}, for Ly$\alpha$, \civ, \heii\ and \ciii, respectively. Radio-loud quasars are marked with open circles.

We note that the Ly$\alpha$ halo morphologies and surface brightness are consistent with those obtained by \cite{borisova16} and by \cite{fabrizio19}, despite that slightly different method of data processing and maps extraction.

The detection rate of Ly$\alpha$ nebulae is 100\% for our sample of quasars, which has already been reported by \citet{borisova16} and \citet{fabrizio19}. All 4 objects presented in Figure~\ref{fig:linemap_objs} show enormous and asymmetric Ly$\alpha$ nebulae. It is not clear whether the morphology of these Ly$\alpha$ blobs reflects the distribution of the CGM or the way with which the CGM is illuminated by the (anisotropic) light of the quasar.

In Figure~\ref{fig:linemap_objs}, \civ\ nebulae are more compact than their Ly$\alpha$ counterparts for all four objects. This effect is certainly partly due to the \civ\ line being fainter and therefore its detectability threshold is reached at smaller radii; however, in Section~\ref{subsec:results_averageprofiles} we will show that the \civ\ radial profile is indeed steeper than Ly$\alpha$. In total, extended \civ\ emission is detected for 15 quasars, as shown in \ref{fig:civ_halos}. Among these 15 quasars detected in extended \civ, 4 are radio loud, and 11 are radio quiet.

The \heii\ morphology is generally similar to that of \civ, as is shown in Figure~\ref{fig:linemap_objs}, which is not surprising given that they have comparable ionization potentials. Over all 80 quasars, extended \heii\ emission is detected for ten of them, as shown in Figure~\ref{fig:heii_halos}. Three objects are radio loud, and the others are radio quiet. The \civ\ and \heii\ nebulae span a range of diverse morphologies in the sense that most of them are asymmetric, similar to the shapes of their Ly$\alpha$ nebulae. The examples in Figure~\ref{fig:linemap_objs} show clear filamentary/elongated shapes.

The diffuse \ciii\ emission is much fainter than \civ\ and \heii. None of the four objects shown in Figure~\ref{fig:linemap_objs} is detected with diffuse \ciii\ emission. Within the whole sample, only 4 objects are detected with extended \ciii\ emission, with one of them being radio loud, as shown in Figure~\ref{fig:ciii_halos}. For objects at redshift 3 to 4, the \ciii\ line is redshifted to a wavelength range  affected by  several OH sky lines. Therefore, the \ciii\ data are noisier, and it is difficult to obtain reliable measurements of their nebular morphology.

\nv\ is hardly detected, but at this wavelength there are more problems in properly removing the contribution from the strong and broad profile of the Ly$\alpha$ from the Broad Line Region (BLR) of the quasar and also the underlying continuum, which has a discontinuity around Ly$\alpha$ due to IGM absorption. These are obviously issues also for Ly$\alpha$, but are more relevant for \nv\ as it is much fainter than Ly$\alpha$.
Anyhow, as discussed in \citet{nagao06}, the diagnostics involving nitrogen are more difficult to interpret as they rely on the assumed nitrogen abundance scaling with respect to other elements, which is poorly known in AGNs and especially at high redshift \cite[see][and references therein]{dors17,dors19}.

\subsection{Line Ratios Diagnostics and Metallicity}

We use the \civ, \heii, and \ciii\ emission to estimate the metallicity of the CGM gas based on  AGN photoionization models. \civ, \heii, and \ciii\ are lines whose ratios have been commonly applied to estimate the metallicity and ionization coditions in radio galaxies, AGNs and star forming galaxies \cite[e.g.][]{nagao06, dors14, dors19, gutkin16, montero17, matsuoka18, nakajima18, mignoli19}.

\begin{figure}[ht!]
\includegraphics[width=0.5\textwidth]{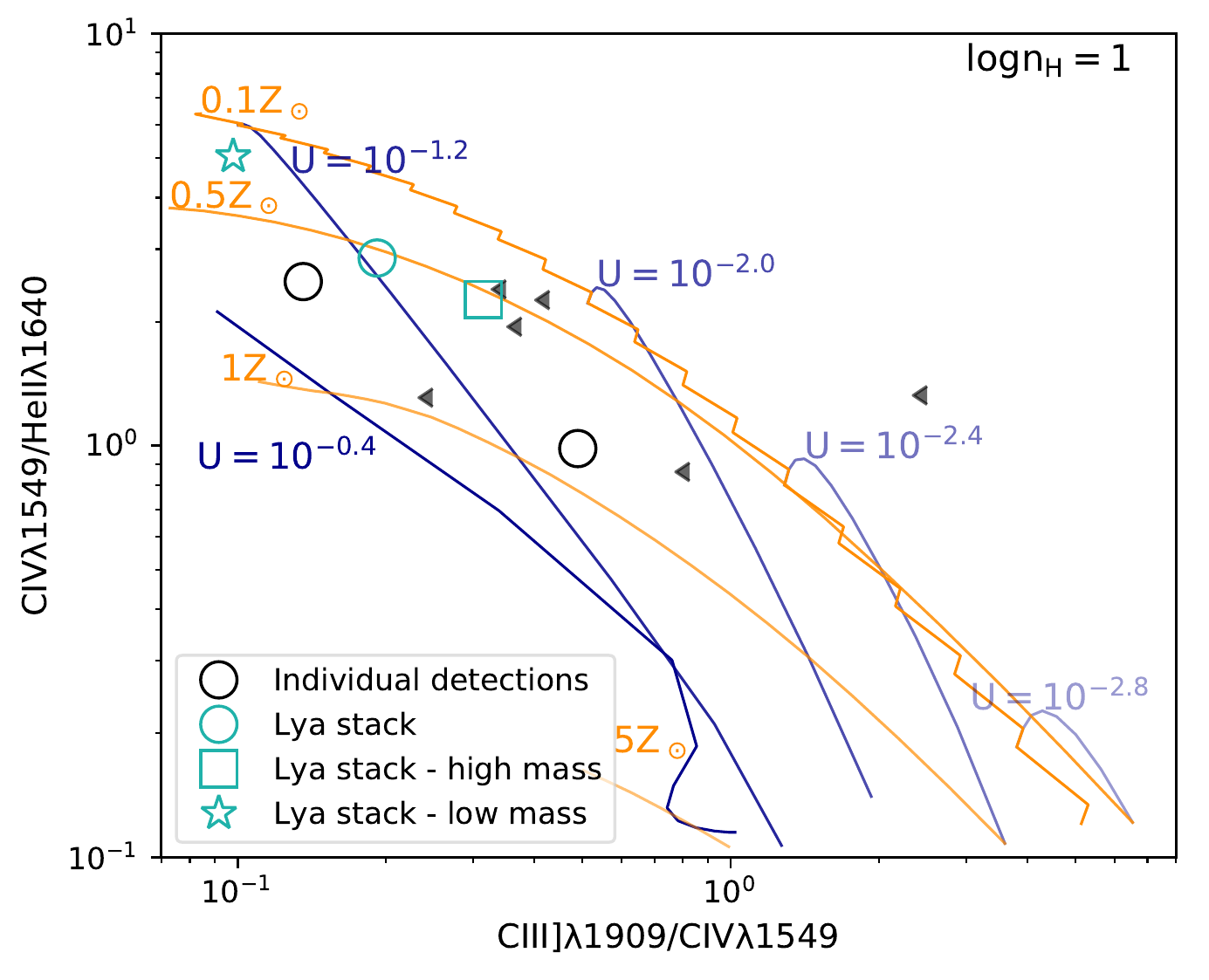}
\caption{\civ/\heii\ vs \ciii/\civ\ diagnostic diagrams. Lines show the results of quasar photoionization models at fixed ionization paramenter (blue) and at fixed metallicity (orange). Black open circles show the line ratios of the only two quasars with detection of exteded emission for all three lines, \civ, \heii, and \ciii. Black triangles show the upper limits of objects with \civ\ and \heii\ detections, but without \ciii\ detections. The light blue circle shows the result of stacking all spaxels with Ly$\alpha$ detection. The results obtained by splitting the sample in high-mass and low-mass BHs are shown with a square and a star, respectively.\label{fig:diagnostic_line}}
\end{figure}

\begin{figure*}[ht!]
\plotone{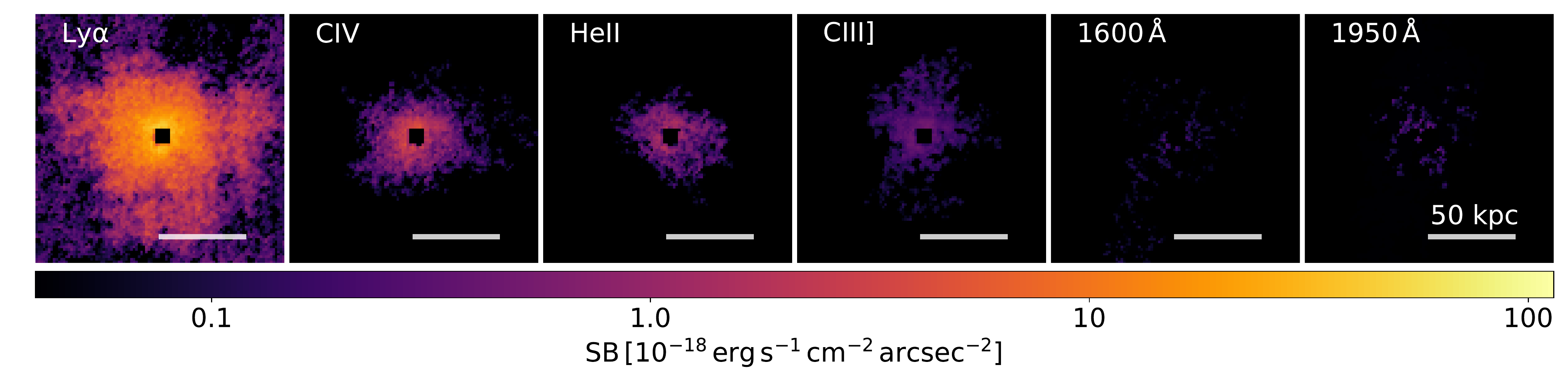}
\caption{Maps of Ly$\alpha$, \civ, \heii\ and \ciii\ resulting from the stacking of all MUSE cubes. We also plot the residuals of the continuum emission at 1600\AA\ and 1950\AA\ for comparison. Each thumbnail has the same size as in Figure.~\ref{fig:linemap_objs}. \label{fig:stack1}}
\end{figure*}

We first use the models provided by \citet{nagao06} and \citet{matsuoka18}. These models are mainly based on the CLOUDY models, using a typical quasar-like ionizing continuum, spanning a wide range of metallicities and ionization parameters. Figure~\ref{fig:diagnostic_line} shows the grid of ionization models for a range of gas metallicities and ionization parameters.
The blue and orange lines show models at constant ionization parameter and at constant metallicity, respectively.
These models are obtained assuming a gas density $\mathrm{n_\mathrm{H}\, =\, 10\, cm^{-3}}$,
which is appropriate for the clumpy gas on large (kpc) scales in the circumgalactic medium which is responsible for the bulk of the nebular line emission under investigation; we will further support this choice (and show that it is a conservative assumption) through the guidance of cosmological simulations as discussed in Section.~\ref{subsec:simulations}. Typically, the inner Narrow Line Region (NLR) of AGNs have even higher densities \citep[e.g.][]{mingozzi19}, which would result in metallicities even higher than estimatated in the following \citep{nagao06,matsuoka09}.
Recent spatially resolved observational studies  \citep[e.g.][]{revalski18}
have shown that the electron density in some NLRs decreases with the radius.
However, \cite{dors19} found that electron density variations as a function of distance from the AGN
have an almost negligible influence on the predicted line ratios and,
consequently, on metallicity estimations.

The temperature of the gas in these models is between $7 \times 10^3$ and $2 \times 10^4$~K, as for most photoionization models, whereby the temperature is kept in this range by the thermostat effect of various nebular cooling lines.

We only have two objects with detections of all three emission lines. They are shown as the open circles in Figure~\ref{fig:diagnostic_line}. Their line ratios fall in the region with a metallicity $0.5 Z_\odot<Z< Z_\odot$. For the objects with \civ\ and \heii\ detections but without \ciii\ detections, we provide the upper limits of the \ciii/\civ\ ratios, as shown by the black triangles. These result into metallicity lower limits in the region of $0.1Z_\odot<Z<1Z_\odot$. It should be noted that the line flux for these individual objects is measured within a $3\arcsec$ aperture, which traces a relatively central region of about 8--24 kpc.

In order to constrain the average metallicity of Ly$\alpha$ nebulae we stack together all spaxels with detection of Ly$\alpha$ from all 80 objects; this provides an average flux of the other UV lines in the resulting stacked spectrum. The ``S/N clipped'' datacubes produced in Section~\ref{subsec:data_3dextract} allows us to easily pick out spaxels with detection of Ly$\alpha$.
To examine possible trends, we also split the sample based on the black hole mass of the individual quasar, computed from the quasar continuum luminosity at 1450\AA\ and FWHM of \civ, following the virial method given in \citet{trakhtenbrot12}. Since most of the nebulae are faint, we simply divide the sample into two halves.
The inferred black hole masses for individual objects are listed in Table~\ref{tab:sample}. The median value of black hole mass is $10^{10.26}$\,M$_\odot$. The line ratios resulting from stacking all 80 objects is shown with the light blue circle in Figure~\ref{fig:diagnostic_line}. The high-mass and low-mass subsamples are indicated with a square and a star, respectively.

The average metallicity for the Ly$\alpha$ nebulae resulting from the stacking of the whole sample is about $0.5Z_\odot$, smaller than the two individual detections shown by black circles. The properties of high-mass and low-mass subsamples are quite similar, with the low-mass subsample having slightly lower metallicity, which is likely mirroring the mass-metallicity relation at these redshifts \citep{maiolino08,troncoso14,onodera16} extending also to quasars \citep{matsuoka11,xu18} and on large scales \citep[but see][]{mignoli19}.

We have also interpreted the line ratios by using the photoionization models presented by \cite{dors19}, which derived two semi-empirical calibrations between the metallicity of the NLR of type-2 AGN and the rest frame of the \nv/\heii, C43 = log[(\civ\ +\ciii)/\heii], and \ciii/\civ\ emission-line intensity ratios. We use their C43 versus \ciii/\civ\ diagram to derive the  metallicities of the CGM in our sample. With those models we obtain even higher, super-solar metallicities for the two objects with detections of all three emission lines. Specifically, we find metallicities in the range 2--4~$Z_{\odot}$ and ionization parameter $\log{U}$ in the range between $-1.5$ and $-1.0$.

\begin{figure}[ht!]
\includegraphics[width=0.4\textwidth]{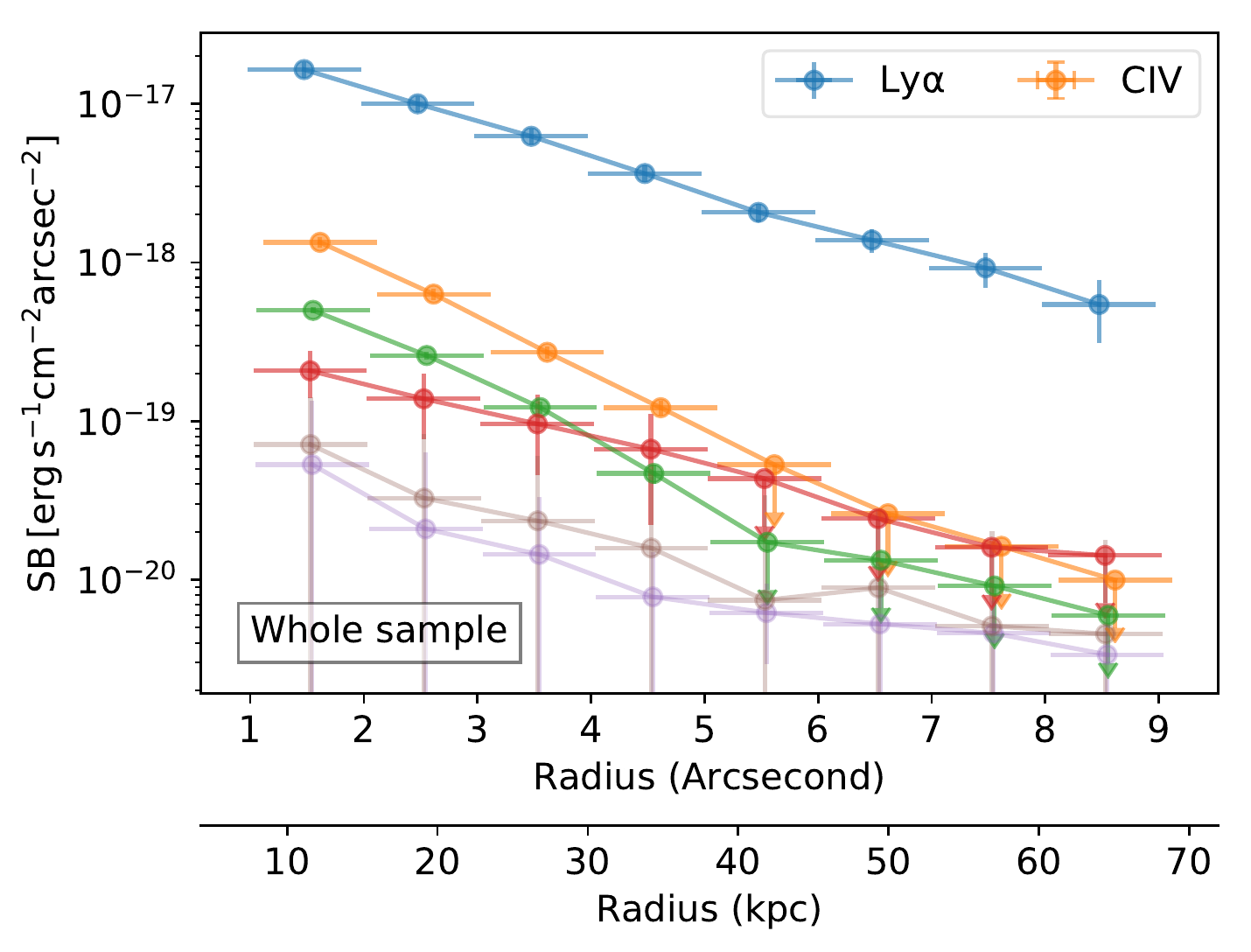}
\includegraphics[width=0.4\textwidth]{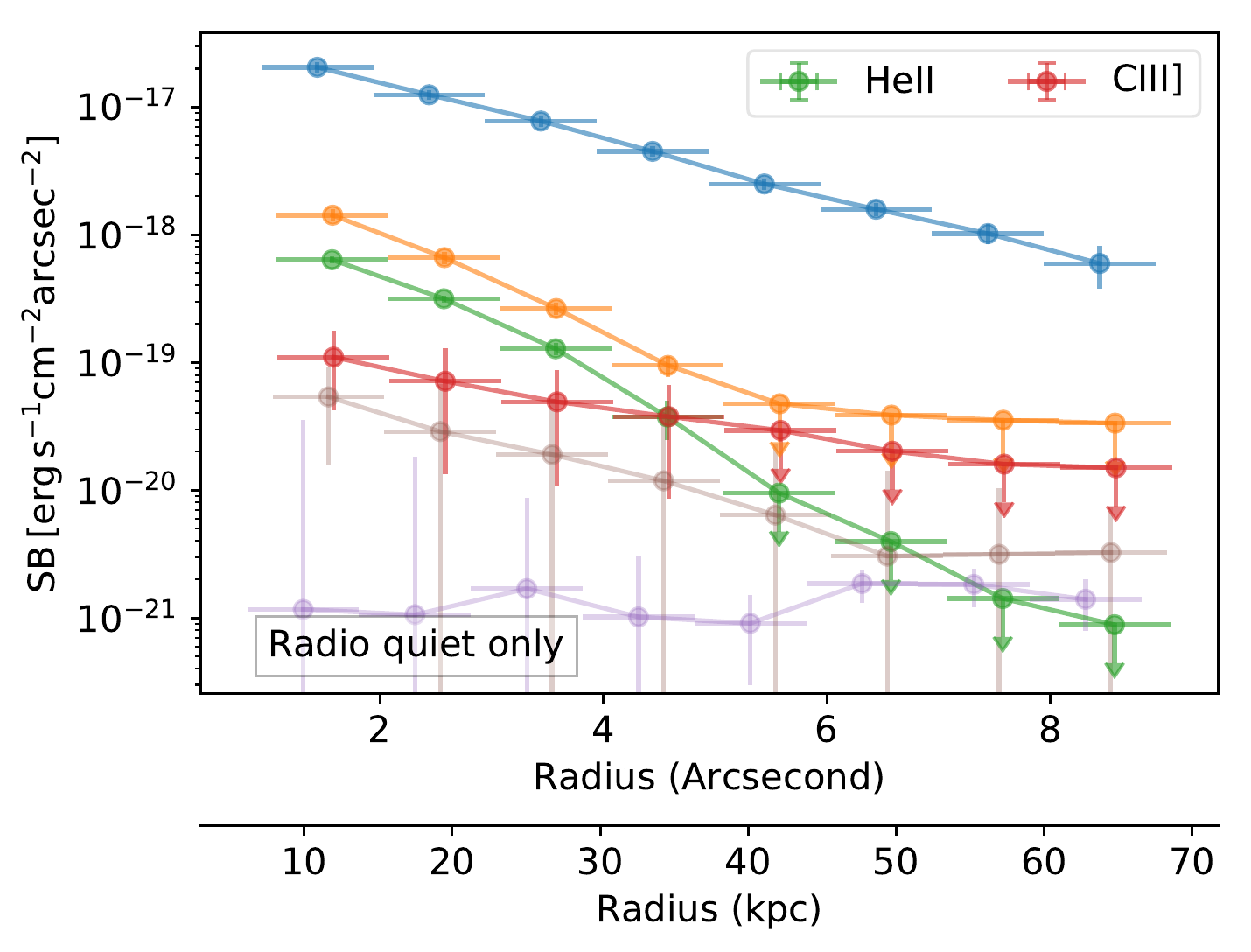}
\includegraphics[width=0.4\textwidth]{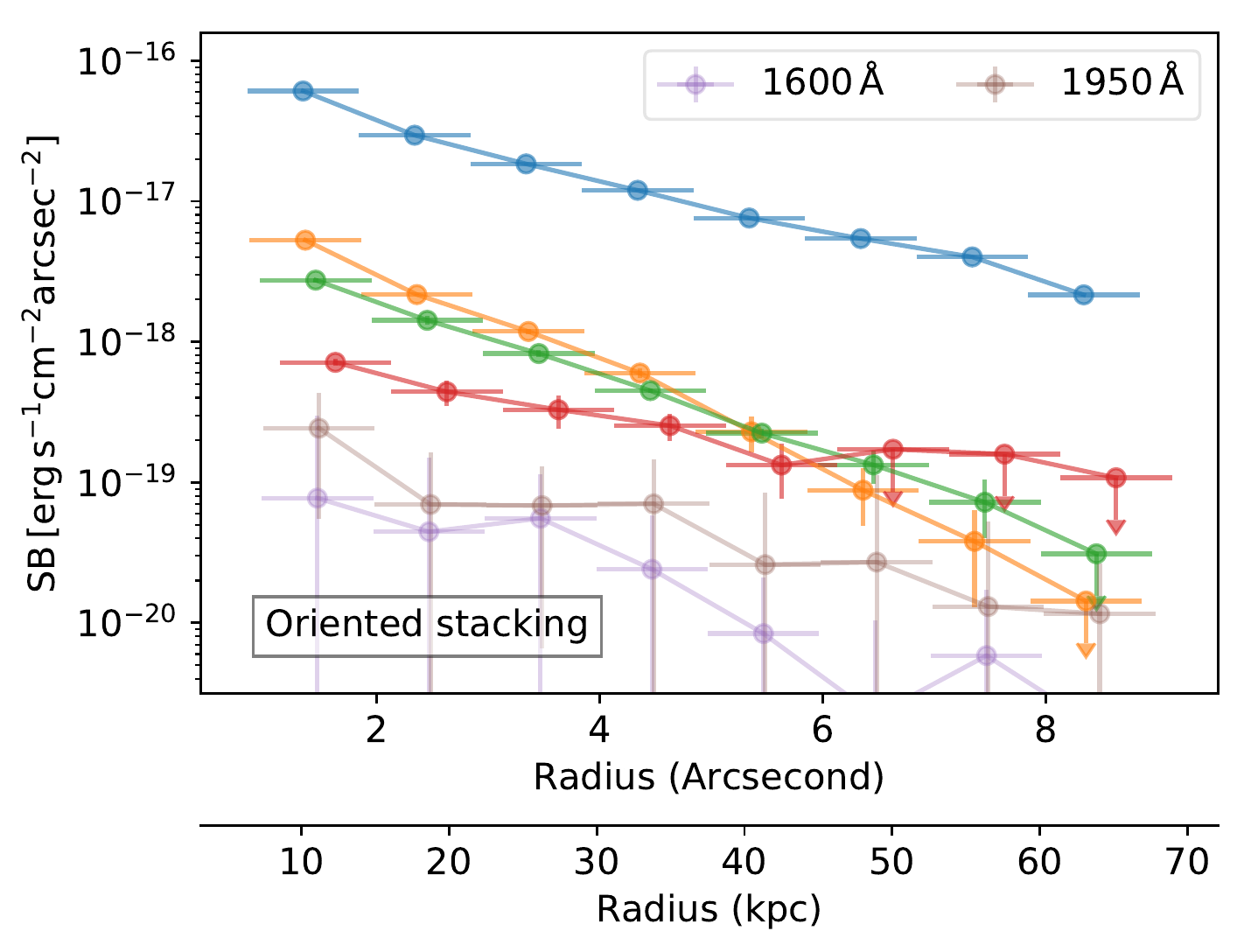}
\caption{Average surface brightness profiles of the nebular lines for the stacked datacubes.
For comparison, the radial profiles of PSF continuum subtraction residual at 1600\AA\ and 1950\AA\ are also shown. The top panel shows the result of stacking all cubes of all quasars. The central panel shows the result of stacking only radio quiet quasars. The bottom panel shows the profile extracted along a  pseudo-long slit on the stack obtained after re-aligning the cubes along the Ly$\alpha$ extension. In each panel, the bottom scale gives the radius in kpc at the median redshift of the sample. \label{fig:stack1_profile}}
\end{figure}

\begin{figure}[ht!]
\includegraphics[width=0.4\textwidth]{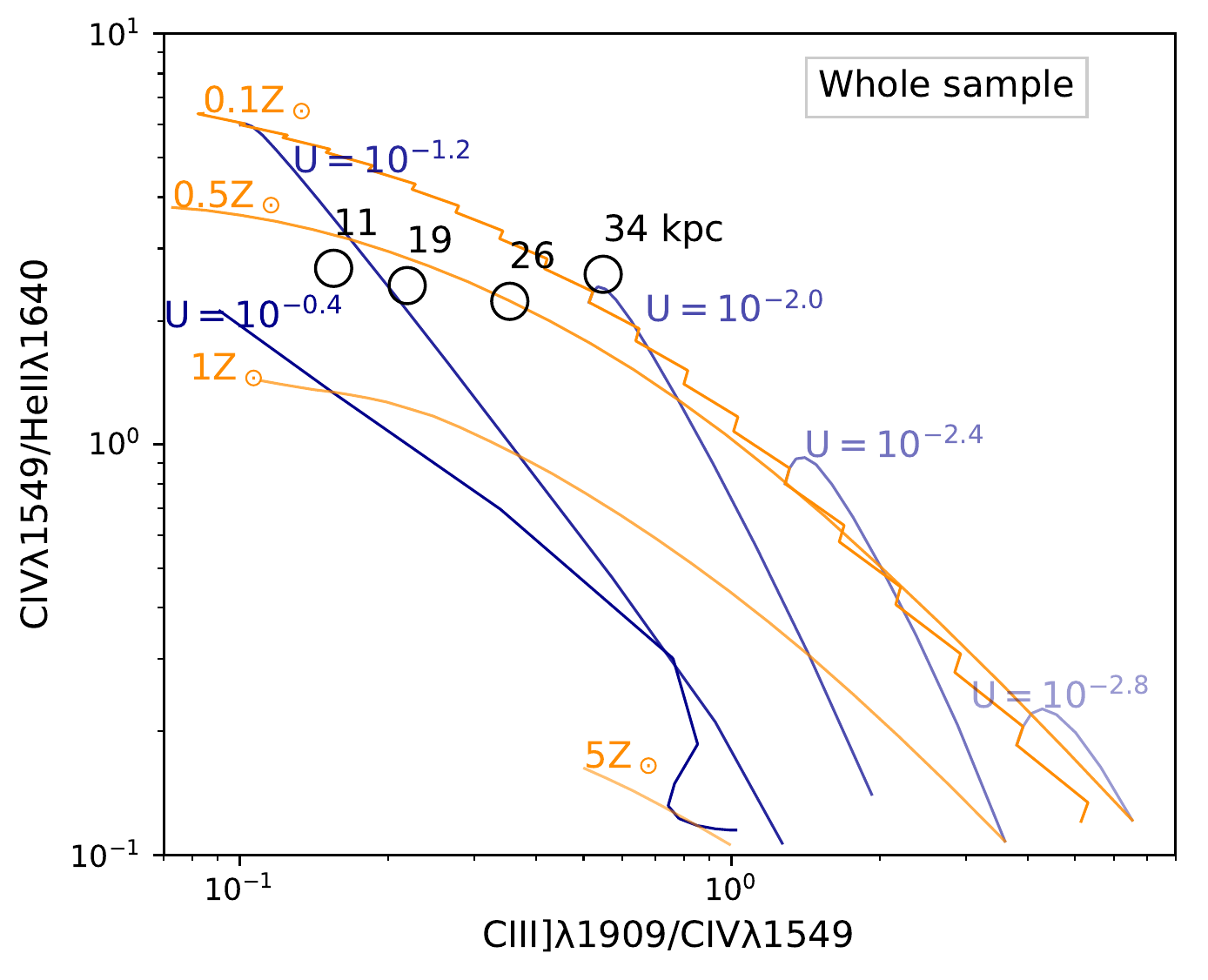}
\includegraphics[width=0.4\textwidth]{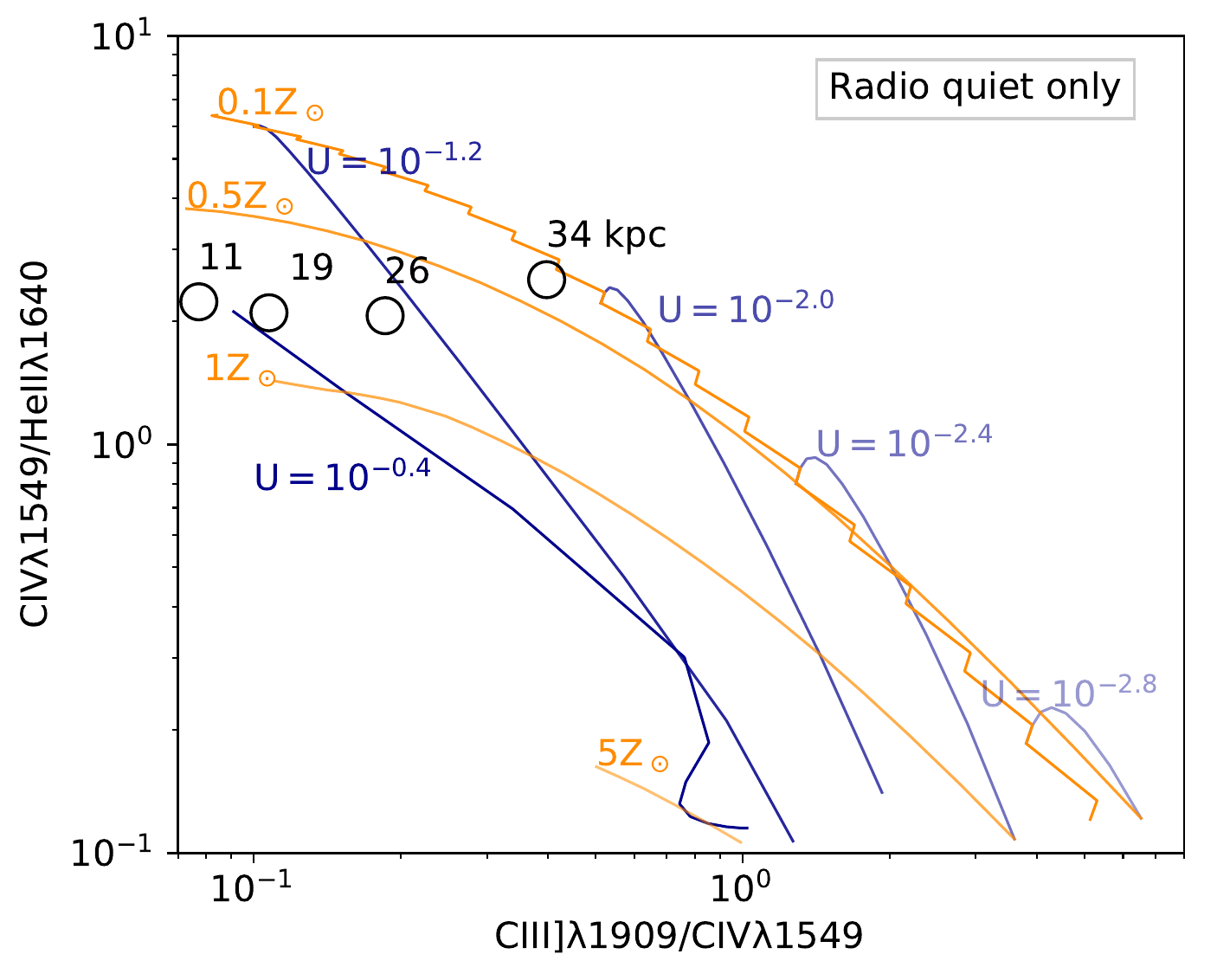}
\includegraphics[width=0.4\textwidth]{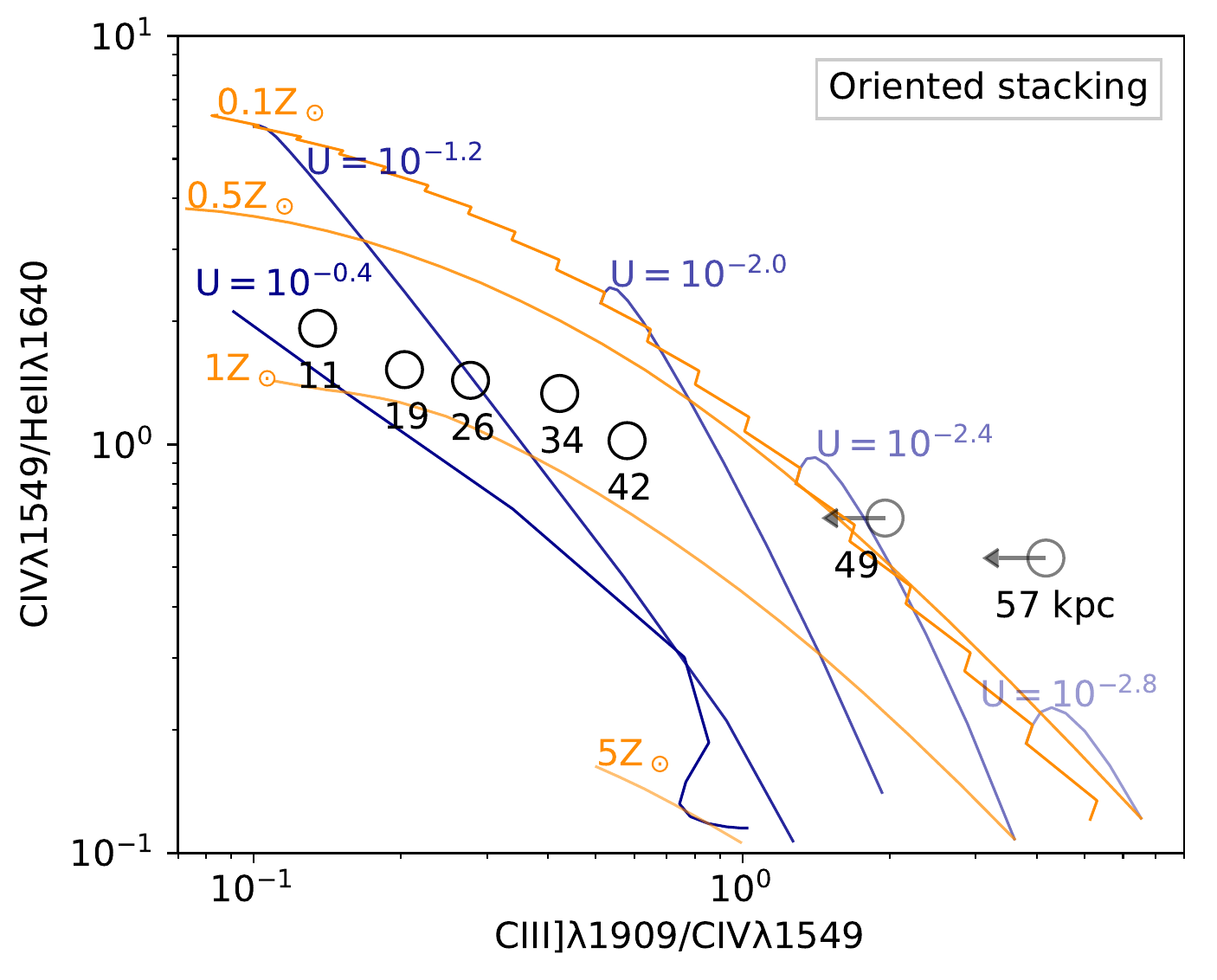}
\caption{The same diagnostic diagram as in Figure~\ref{fig:diagnostic_line} for the line ratios inferred from the stacked cubes. The open circles in different colors show the line ratios at different distances from the center. The distance is given in kpc next to each symbol.
The top panel shows the result from the stacking of all cubes of all quasars. The central panel shows the result of stacking only radio quiet quasars. The bottom panel shows the line ratios extracted from the pseudo-long slit one the stack obtained after re-aligning the cubes along the Ly$\alpha$ extension.\label{fig:stack1_metallicity}}
\end{figure}

\subsection{Cubes stacking and Average Radial Profiles}
Despite the deep observations of the sample, we can only directly detect the extended \civ, \heii, and \ciii\ emission lines in a small fraction of the quasars. To increase the S/N on faint lines, determine the average line strengths and obtain spatially resolved information on the properties of the CGM, we adopt a full 3D-cubes stacking procedure. We try two different stacking startegies: 1) stack all objects with their (random) orientation on sky, 2) stack datacubes after re-aligning them along their Ly$\alpha$ extensions. We apply this procedure to the datacubes after the PSF and continuum subtraction.

\subsubsection{Stacking with Random Orientation} \label{subsec:results_averageprofiles}

Firstly, we try to stack all objects with random orientation.
We shift individual datacubes and re-bin them to a common (rest-frame) wavelength frame. The cubes are also re-aligned spatially to the position of the quasar continuum position. The cubes are averaged and weighted by their exposure time. We visually inspect each datacube to avoid contamination from bright stars and other artifacts in the field of view. We have also stacked the cubes by taking their median and the result is almost indistinguishable from the average-stacked one. Finally, we dynamically extract the stacked nebular line emission maps following the procedure described in Section~\ref{subsec:data_3dextract}, i.e. extracting pseudo-NB images after masking each pixel in the cube by the same S/N threshold given in Section~\ref{subsec:data_3dextract}.

The resulting maps of Ly$\alpha$, \civ, \heii, and \ciii\ are shown in Figure~\ref{fig:stack1}. A clear, large average Ly$\alpha$ nebula is visible, and fainter but clear \civ, \heii\, and \ciii\ nebulae are also seen.

\begin{figure*}[ht!]
\plotone{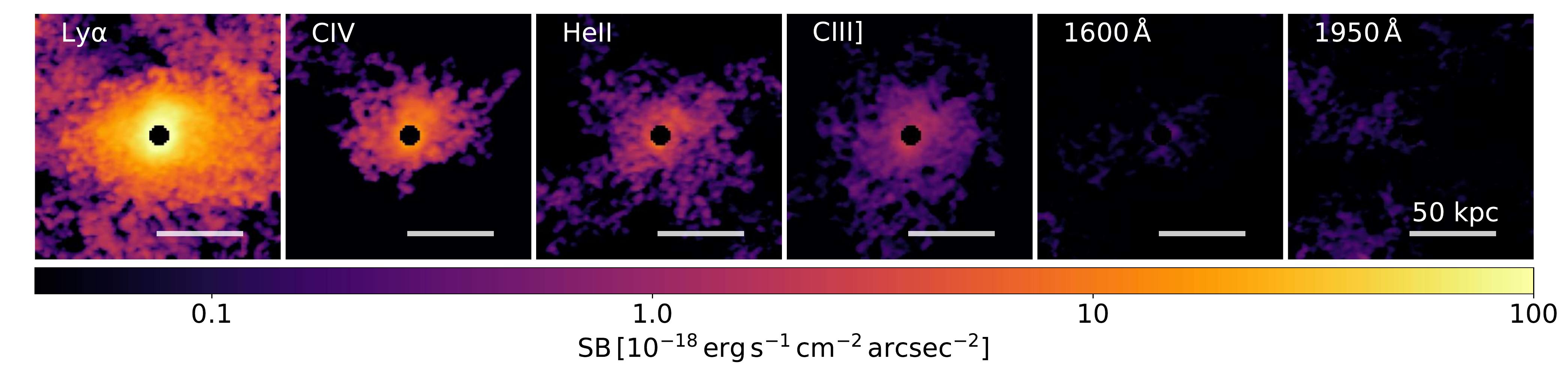}
\caption{Maps of Ly$\alpha$, \civ, \heii, \ciii, and residuals at 1600\AA\ and 1950\AA, obtained from the cubes stacked after re-aligning them along the primary Ly$\alpha$ extension.  Each thumbnail has the same size as in Figure.~\ref{fig:linemap_objs}.\label{fig:stack_rotate}}
\end{figure*}

\begin{figure*}[ht!]
\plotone{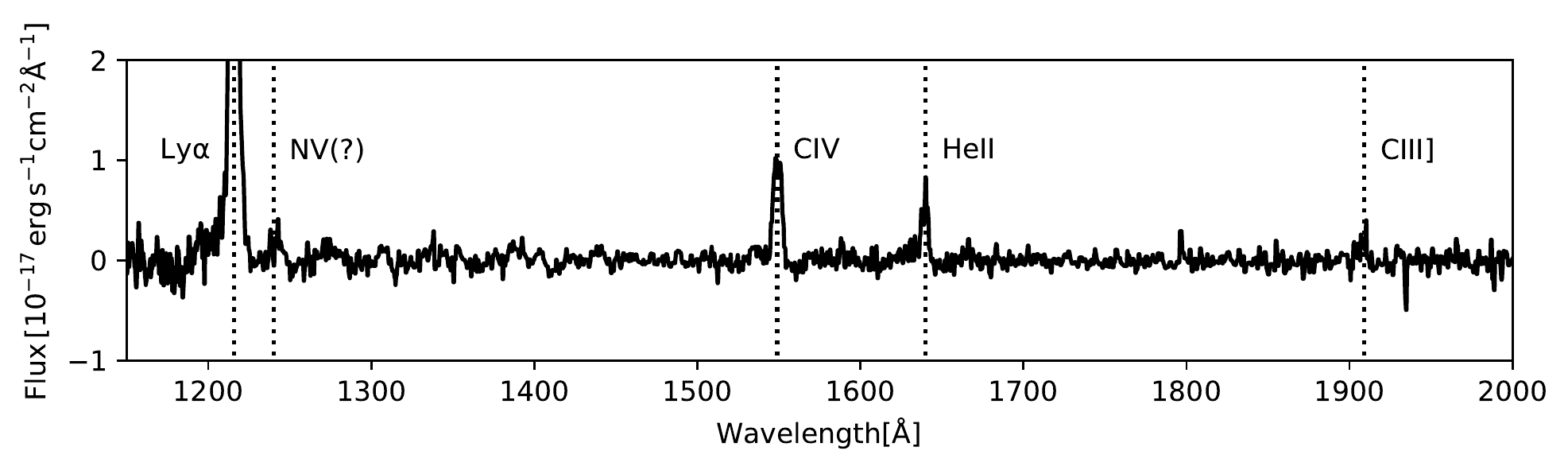}
\caption{The 1D spectrum extracted by a pseudo long slit on the cube obtained by staking the individual cubes along the primary Ly$\alpha$ extension.\label{fig:stack_rotate_spe}}
\end{figure*}

We measure average radial profiles from these line maps by extracting the flux averaged in radial annuli. The top panel Figure~\ref{fig:stack1_profile} shows the average radial surface brightness profile for Ly$\alpha$, \civ, \heii, and \ciii\ in blue, orange, green, and red, respectively. We also plot the radial profile of PSF-subtracted continuum residuals in the wavelength range between \civ\ and \heii\ (i.e. 1600\AA) and at a wavelength just next to CIII] (specifically at 1950\AA). Since there are no prominent nebular lines at these wavelengths and the continuum extended emission should be negligible, the emission at these wavelengths is likely only associated with residuals from the PSF subtraction or from residual sky emission (in particular residual, unmasked OH emission lines) and therefore gives us a measure of the reliability of the detection of the nebular lines.

We note that the radial profile of the Ly$\alpha$ emission resulting from the
stacked cube is consistent with the average radial profile of Ly$\alpha$ obtained by
\cite{borisova16} and by \cite{fabrizio19}, despite the fact that they adopt a different methodology (they
average the radial profiles of the individual objects while we extract the radial profile of the stacked cube).

As illustrated in the top panel of Figure~\ref{fig:stack1_profile}, the Ly$\alpha$ surface brightness is about one order of magnitude higher than that of \civ, \heii, and \ciii, and the gap is even larger at larger radii. The average profile of \civ\ and \heii\ is also about one order of magnitude brighter than the PSF subtraction residual at 1600\AA. This means our results are reliable. The \ciii\ radial profile is closer to the 1950\AA\ continuum residual radial profile, indicating that the detection of this line is indeed more marginal, at the 2$\sigma$ level (the detection will become more significant with the approach discussed in the next section when aligning the nebulae). Beyond a radius of $\sim$40 kpc none of the emission lines, except for Ly$\alpha$, is detected.

The \civ\ radial profile is steeper than the Ly$\alpha$ profile, which could be naively interpreted as radially decreasing metallicity. However, the \heii\ has the same steep radial profile as \civ, which suggests that the decrease of these two lines is more a consequence of a decreasing ionization parameter as a function of radius (which is expected if the gas density decreases more slowly than r$^{-2}$). The \ciii\ line, despite being more marginally detected, decreases less steeply than \civ\ and \heii, and follows a similar decline as Ly$\alpha$. This reinforces the idea that the radial variation of metallicity is not strong.

We also repeat the stacking and profile extraction excluding the radio-loud quasars.
As it has been claimed that metal lines in the CGM are perhaps associated with (rare) radio jets \citep{borisova16}, by excluding those few radio loud quasars in the sample we can assess if the result is potentially dominated by the few radio loud sources. The resulting nebular lines profiles obtained by stacking only the radio-quiet quasars are shown in the central
panel of Figure~\ref{fig:stack1_profile}. There are no significant differences with respect to the result obtained by stacking
the whole sample (top panel). The main difference is for the upper limits at large radii that, in the case of  staking only the radio-quiet quasars, are less constraining.

Figure~\ref{fig:stack1_metallicity} shows the diagnostic line ratios at different radii resulting from the cube staking. The line ratios at different distances are shown by colored circles and the physical distances are labelled next to each circle. The top panel shows the
result from stacking the whole sample while the central panel is the result from stacking only the radio-quiet quasars.
The metallicity remains stable at about $0.5Z_\odot$ out to $<$30 kpc, while the ionization parameter decreases, as expected. The diagnostics measured at 34~kpc suggest that both metallicity and ionization parameter decrease at even larger radii.

\subsubsection{Stacking along Ly$\alpha$ Extensions}

In Section~\ref{subsec:results_halos}
we have shown that most strong Ly$\alpha$ nebulae are not radially symmetric and that the shapes of their \civ, \heii, and \ciii\ nebulae are, in general, similar to their Ly$\alpha$ counterparts.
Therefore, a random stacking of datacubes is not an efficient way to enhance the S/N at large distances. Moreover, the random stacking may result into a biased result, in the sense that the presence of clumps with high metallicity and high surface brightness may be washed out by the random averaging method.
In this section we stack the nebular emission by re-aligning them along the direction of the primary Ly$\alpha$ extension.

We select 15 objects that have very extended and asymmetric Ly$\alpha$ nebulae (IDs 8, 10, 13, 18, 21, 33, 47, 49, 50, 51, 56, 65, 67, 74, 75). Note that in this case we also include for completeness 4 radio loud quasars, as they are among those showing asymmetric extended nebulae, but the results do not change significantly by excluding these four objects.
We find the major axis for each target according to the shape of Ly$\alpha$ nebulae via a principal component analysis (PCA). For each Ly$\alpha$ image, we calculate its covariance matrix, eigenvectors, and eigenvalues. Then we align each datacube horizontally based on the major axis via a linear transformation. The rotating transformation matrix is calculated by the major eigenvector.
We then proceed with the same stacking procedure discussed in the previous section using the re-aligned cubes.

Figure~\ref{fig:stack_rotate} shows the line maps resulting from the re-aligned stacked cubes. In each map, the surface brightness distribution is obviously stretched in the horizontal direction, especially in Ly$\alpha$. Note that the outer boundaries of the nebulae are now more irregular and noisier than in the maps from the global stacking as a consequence of the much smaller number of haloes being combined here. We apply a horizontal pseudo-long slit with a width of 20 spaxels to the stacked datacube. We only use the data in this slit and avoid other areas with little signal (which would contribute primarily only with noise and sky residual emission). Using this method we cannot obtain a truly average radial surface brightness distribution, but the average properties for such regions along  the extended Ly$\alpha$ emission. The 1D spectrum extracted from such a pseudo-slit are shown in Figure~\ref{fig:stack_rotate_spe}, which clearly reveals the detection of \civ\ and \heii. \ciii\ is more marginally detected, but significant, as discussed in the following.
\nv\ may also be detected, but as already mentioned in this spectral region potential systematic residuals from the strong broad Ly$\alpha$ emitted by the quasar BLR makes the spectral features uncertain and questionable.

The nebular lines radial profiles along the Ly$\alpha$ extension are shown in the bottom panel of Figure~\ref{fig:stack1_profile}.
As expected, the surface brightness of all nebular lines is significantly enhanced  compared to the average results shown in Figure~\ref{fig:stack1_profile}. The S/N is higher, including for \ciii, which is better detected. The significance of the detection for all lines is confirmed by the comparison with the radial profiles of the residuals extracted at 1600\AA\ (for \civ\ and \heii\ ) and at 1950\AA \ (for \ciii ). The detections now extend to larger distances of about $5 \arcsec$ (corresponding to about 45 kpc at the average distance of these quasars).

Also in this case \civ\ and \heii\ decrease more steeply than Ly$\alpha$ while \ciii\ shows a flatter profile, again supporting the idea that the bulk of the decrease of the \civ\ and \heii\ is primarily due to the ionization parameter, while the metallicity does not change dramatically.

We derive the metallicity and ionization parameter of the gas along the Ly$\alpha$ extension in the bottom panel of Figure~\ref{fig:stack1_metallicity}. The metallicity of these  regions stays stable at $0.5Z_\odot<Z<Z_\odot$ out to about 42 kpc, but the ionization parameter decreases with the distance, as expected. It is not surprising that the metallicity in this case is higher than in the randomly oriented stacking. Indeed, as already mentioned, along the Ly$\alpha$ extension we are probably tracing
gas that has been enriched more recently by the quasar outflow, as quasar-driven outflows and direction of quasar illumination are
generally in the same direction.

\begin{figure*}[ht!]
\includegraphics[width=0.496\textwidth, height=11.5cm]{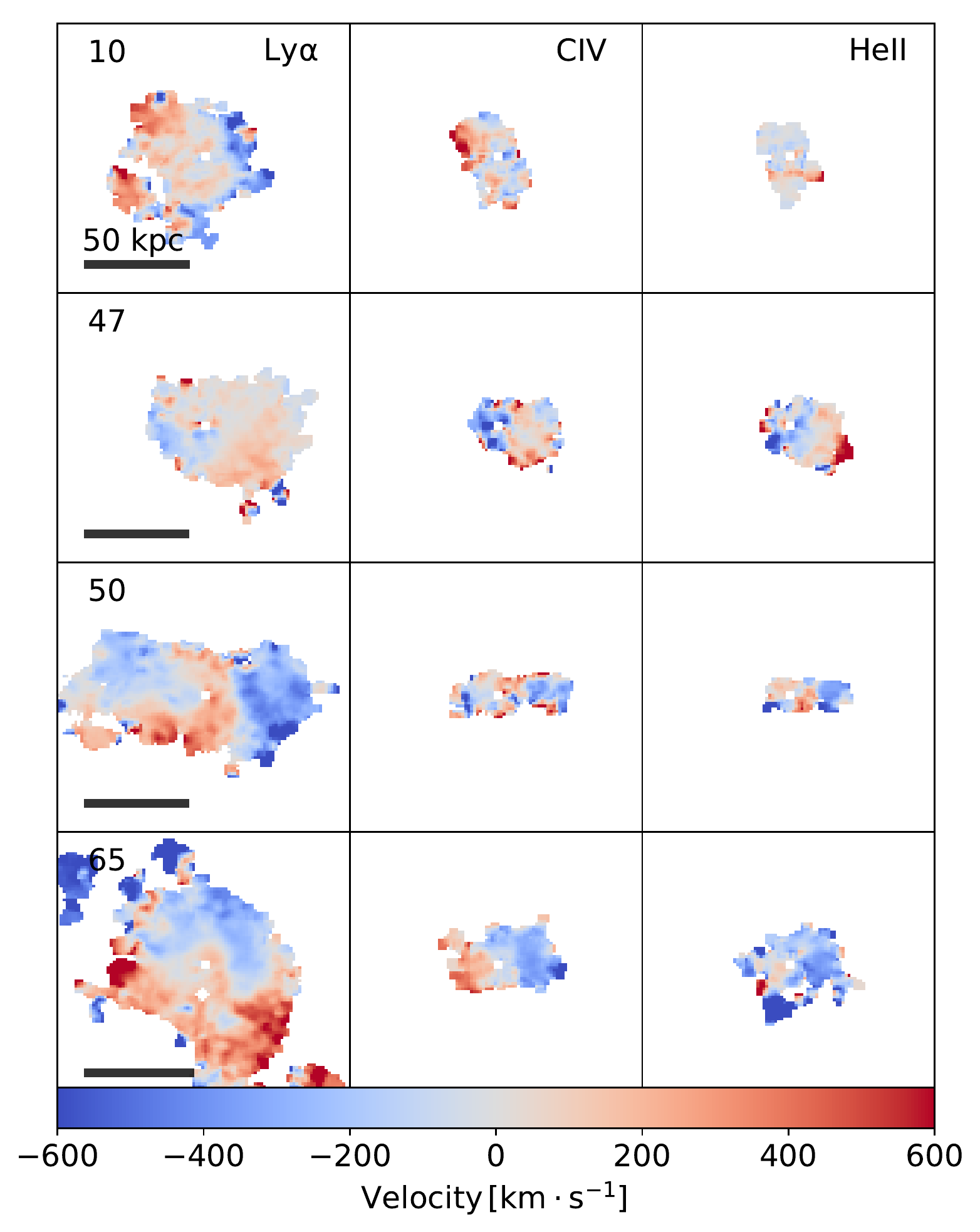}
\includegraphics[width=0.495\textwidth, height=11.5cm]{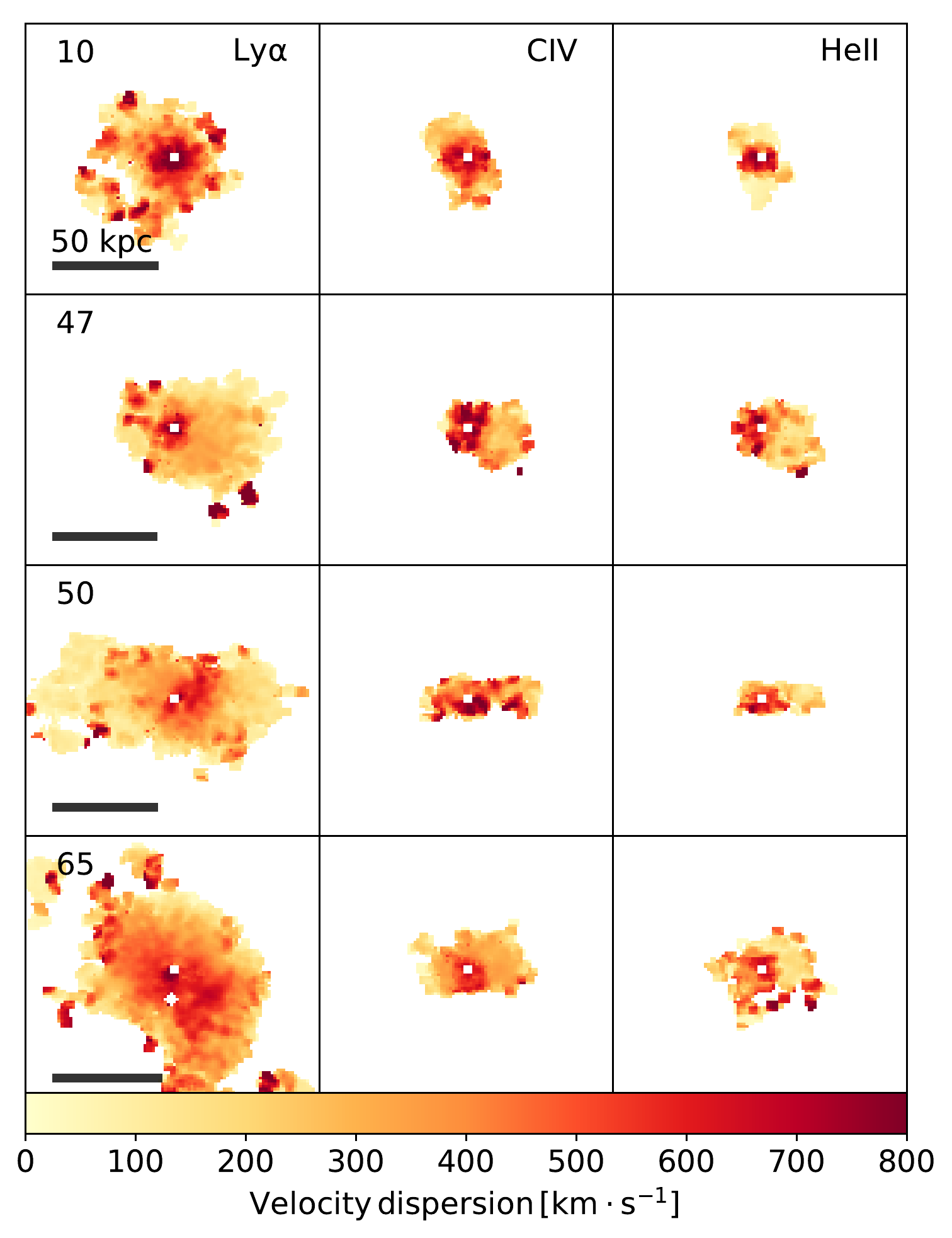}
\caption{Projected velocity field (left panel) and velocity dispersion (right panel) traced by Ly$\alpha$ (left column), \civ\ (middle column), and \heii\ (right column) of the same four objects shown in Figure.~\ref{fig:linemap_objs}. The velocity and velocity dispersion are estimated from the first and second momentum of the flux distribution, respectively. Each thumbnail has the same size as Figure.~\ref{fig:linemap_objs}. \label{fig:velocity}}
\end{figure*}

\subsection{Kinematics}
\label{subsec:results_kinematicss}

We use the ``S/N clipped" datacube defined in Section~\ref{subsec:data_3dextract} to produce the two dimensional maps of flux-weighted velocity centroid and flux-weighted velocity dispersion. We do not fit any emission line profile, but produce the first and second momentum of the flux distribution in the wavelength domain. The first and second momentum maps then indicate the velocity and velocity dispersion, respectively. The velocity and velocity dispersion maps of Ly$\alpha$ nebulae have been published in \citet{borisova16} and \citet{fabrizio19}. In this section, we therefore focus on the kinematics of \civ\ and \heii, and only on the four large nebulae in which the kinematics can be fully resolved with a good S/N (ID 10, 47, 50, 65). \ciii\ is too noisy and barely resolved in any individual object to have the kinematics properly mapped.

Figure~\ref{fig:velocity} shows the resulting velocity field and velocity dispersion inferred from Ly$\alpha$, \civ, and \heii.
As discussed in Section~\ref{subsec:results_halos}, \civ\ and \heii\ are much fainter and more compact than Ly$\alpha$, only tracing the central region. Ly$\alpha$ traces large scale motions that are difficult to identify in terms of simple rotation, inflow or outflows, although there are interesting similarities with what is seen in cosmological simulations, as discussed later on. However, here it is interesting to note that \civ\ and \heii\ broadly share the same velocity pattern as Ly$\alpha$. The additional interesting aspect is that all three nebular lines are characterized by very large velocity dispersions in the innermost region (r$\,<\,2''$), exceeding 700~km/s, likely tracing active outflows, although we can not exclude this may be tracing the virial motions in the central deep gravitational potential well (this will be discussed further when comparing with the numerical simulations).
However, on larger scales the velocity dispersion rapidly drops to much lower values (100--200~km/s), indicating that metal enriched gas on larger scales is dynamically quiescent and likely part of the virialized CGM. A similar kinematic pattern is also observed in a $z\sim5$ quasar \citep[][Ginolfi et al. in prep]{ginolfi18}.

\section{Discussion}
\label{sec:discussion}
\subsection{Implications of High Metallicities in the Circumgalactic Medium}

The metallicity that we infer for the CGM, out to a radius of about 40~kpc in the stacked spectrum (and also in two individual targets) is far too high ($>0.5~Z_{\odot}$) to be ascribed to gas recently accreted from the IGM.
In past works the detection of extended CIV was associated with radio loud quasars \cite[e.g.][]{borisova16} and generally it is typical for giant Ly$\alpha$ nebulae around high redshift radio galaxies that show high \civ\ and \heii\ \cite[e.g.][]{villar07}, which might indicate that the high metallicity gas in these systems has been somehow lifted by the motions associated with the jet. However, in our sample only 4 out of 15 objects that have individual \civ\ detections are radio loud. Moreover, we clearly detected extended metal line emission and high emission also in the stack that excludes radio loud quasars.
\citet{cai17} have found large giant Ly$\alpha$ nebulae with extended \civ\ and \heii\ emission in galaxy overdensities  without any radio detection.
\citet{marino19} have discovered narrow and extended Ly$\alpha$, \civ\ and \heii\ nebulae surrounding a radio quiet quasar at $z=3.02$.
Therefore, there must (also) be mechanisms other than jets promoting the transport of metals to the CGM.

Such high metallicities must have resulted from past extensive enrichment from outflows likely coming from the central galaxy, possibly outflows generated by consecutive quasar phases, which we know are capable of ejecting large amounts of highly enriched gas into the CGM especially in the early Universe \citep[e.g.][]{maiolino12,cano-diaz12,cicone14,cicone15,carniani15,carniani17,bischetti19,fluetsch19}.
Indeed, quasar-driven outflows are fast enough to reach large radii in the halo of the galaxy and  dump enough enriched gas, during consecutive active episodes \citep[see discussion in][]{cicone15}, so that they can bring the metallicity of the CGM to a value close to solar (or even supersolar), within a radius of 50~kpc, already within the first 2 Gyr. Highly metal-loaded SN-driven winds from the central galaxy may also contribute to the enrichment of the CGM on these scales.

An additional source of metals in the CGM could be the enrichment by satellite star forming galaxies, whose ISM may be ram-pressure stripped in the hotter halo of the central galaxy or dispersed through supernova-driven winds.
In the field of view of these objects it is hard to detect any faint satellite galaxies around the quasars, but these may be well below our detection limit and possibly lost in the residuals of the bright quasar light PSF subtraction.

Based on numerical simulations, \citet{hafen19} point out that more than half of the CGM mass originates as (near-pristine) IGM accretion, and wind from the central galaxy is the second important contribution. Gas can be detained in the CGM for billions of years, resulting in a well mixed halo. Therefore, the high metallicity CGM we detect is very likely to trace a mixture of pristine or near-pristine accretion from the IGM, and metal-rich outflows from central galaxies or gas stripped from other satellite galaxies. Our results at least qualitatively support this scenario.

\subsection{Two Components of the CGM} \label{subsec:cgm_components}

According to the simulation of \citet{muratov17}, nearly all metals produced in high redshift galaxies are carried out by galactic winds as far as $0.25R_{\mathrm{vir}}$. At $R_{\mathrm{vir}}$ the outflow metallicity is expected to decrease due to the dilution by the metal-poor component part of the CGM more closely associated with the accreting gas from the IGM. In our results, we do detect a possible metallicity decrease for randomly-oriented stacked results at radii larger than 30~kpc (Figure~\ref{fig:stack1_metallicity}). There seems to be, on average, a metal poor component of the CGM exterior to the central, metal rich one. However, the result is not robust since the signal is extremely weak at large distances.

\begin{figure}[ht!]
\includegraphics[width=0.5\textwidth]{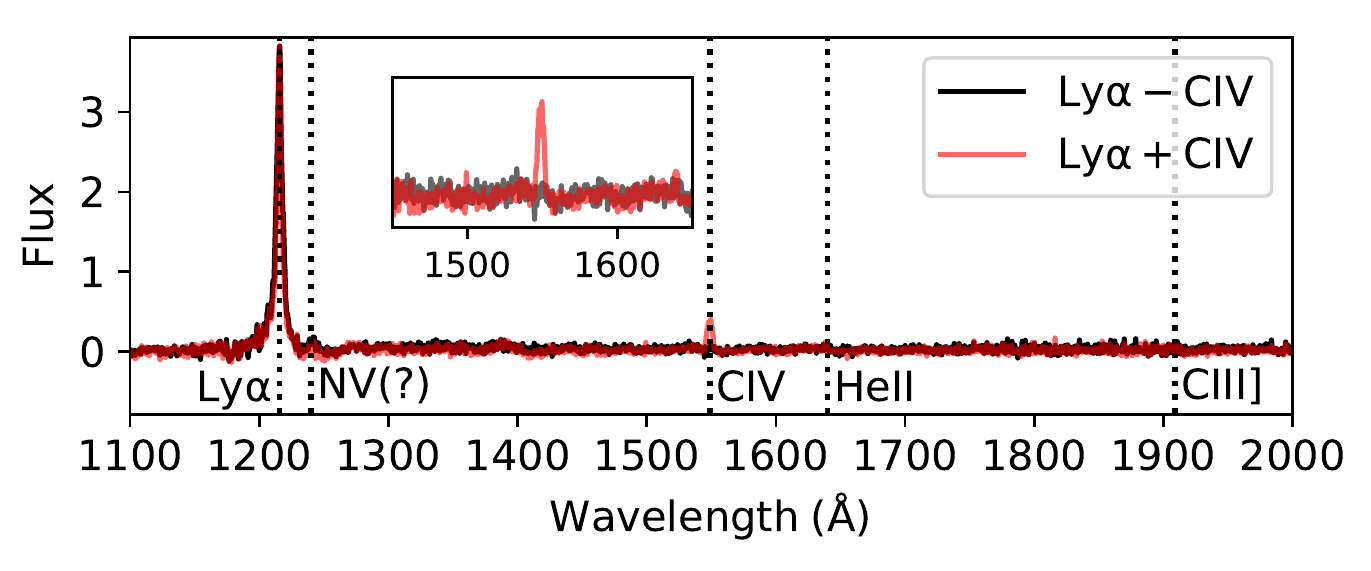}
\caption{Red line: stacking of all spaxels for all objects with both Ly$\alpha$ and \civ\ detections. Black line: stacking of all spaxels with Ly$\alpha$ but without \civ\ detections. The inset shows the zoomed-in spectra around \civ. \label{fig:lya_civ}}
\end{figure}

As discussed in the previous section, the larger extension of the Ly$\alpha$ halo with respect
to the other UV lines is primarily due to the fact that Ly$\alpha$ is much stronger (and also located in a spectral region in which MUSE is more sensitive and with lower sky background) and therefore
can be traced to larger radii from the quasar. The geometrical dilution with radius of the quasar's photoionization radiation (hence the decrease of ionization parameter) also contributes to the decline of high ionization lines such as \civ\ and \heii. However, it is possible that the decrease in metallicity can also contribute to the small extent of \civ\ and \ciii.

Since the metallicity may also vary in a non radially axisymmetric way, because of the complex interplay between near-pristine gas accretion from the IGM, metal-enriched outflows and circulation and mixing in the CGM, we have investigated whether the gas emitting Ly$\alpha$ but not \civ\ (and not \ciii) does show any signature of metal enrichment (i.e. \civ\ emission)
by stacking all spaxels with Ly$\alpha$ but {\it without} \civ\ detection together for all objects. For comparison we also stack all the spaxels {\it with} \civ\ detection (which always also have Ly$\alpha$ detection). Then we scale two spectra to their Ly$\alpha$ peak flux; this is done to investigate the presence or absence of \civ\ relative to the Ly$\alpha$ flux, which is a proxy of the content of ionized gas. The two spectra are shown in Figure~\ref{fig:lya_civ}. The former spectrum is labeled by Ly$\alpha$-\civ, and the latter one is labeled by Ly$\alpha$+\civ. It is quite clear that \civ\ is not detected in the Ly$\alpha$-\civ\ stack and much weaker (below the detection limit), relative to Ly$\alpha$, than in the Ly$\alpha$+\civ \ stack, despite the S/N of the two stacks being similar.
\ciii\ is not detected in both Ly$\alpha$-\civ\ and Ly$\alpha$+\civ\ stack.
This result suggests that the lack of metal lines in the more extended Ly$\alpha$ component of the CGM is likely also associated with a physical lack of metals with respect to the regions detected in CIV.
In other words, there might not be much \civ\ or \ciii\ emission at all outside the area we have already detected with \civ. And this result might hint at a ``hard boundary" between the metal rich component and metal poor component of the CGM. At larger radii Ly$\alpha$ is likely to trace a near pristine component of the CGM recently accreted from the IGM and not yet significantly polluted by the metal enriched outflows produced by the quasar.

We finally note that the Ly$\alpha$-\civ\ stack shows a broad feature near Ly$\alpha$, which may be potentially associated with \nv. However, this feature appears too broad to be really tracing large scale \nv\ emission associated with the CGM. It is more likely that it is a residual of the strong broad component of the Ly$\alpha$. As discussed in the previous sections, \nv\ is more difficult to investigate because of both the subtraction of the bright broad component of the quasar nuclear broad Ly$\alpha$ and the continuum discontinuity in this region.


\begin{figure}[ht!]
\includegraphics[width=0.5\textwidth]{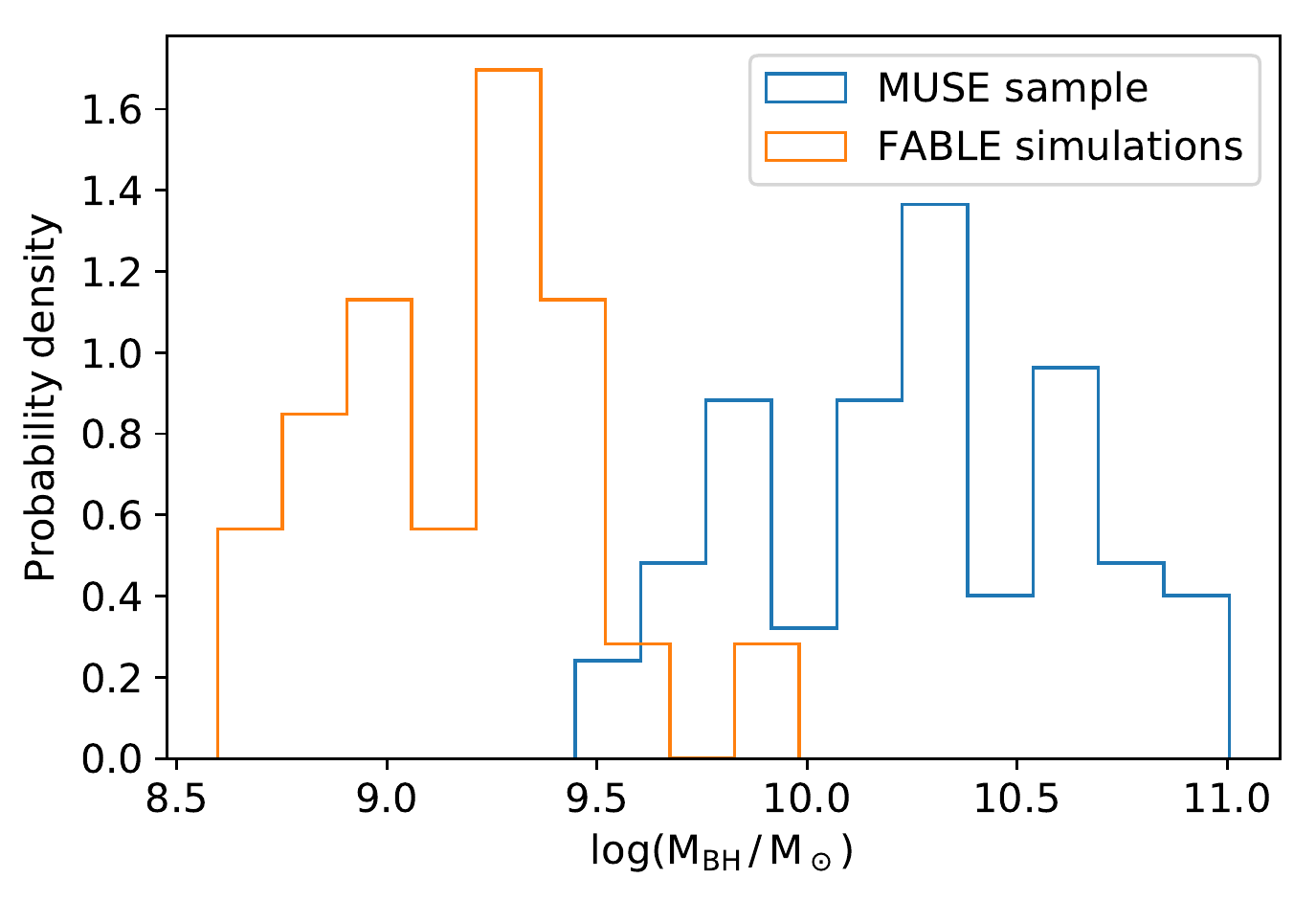}
\caption{Distribution of the black hole masses in the MUSE sample (blue) compared with the distribution of BH masses of the most massive galaxies selected in the FABLE simulation (orange).  \label{fig:BH_mass}}
\end{figure}

\begin{figure*}[ht!]
\includegraphics[width=\textwidth]{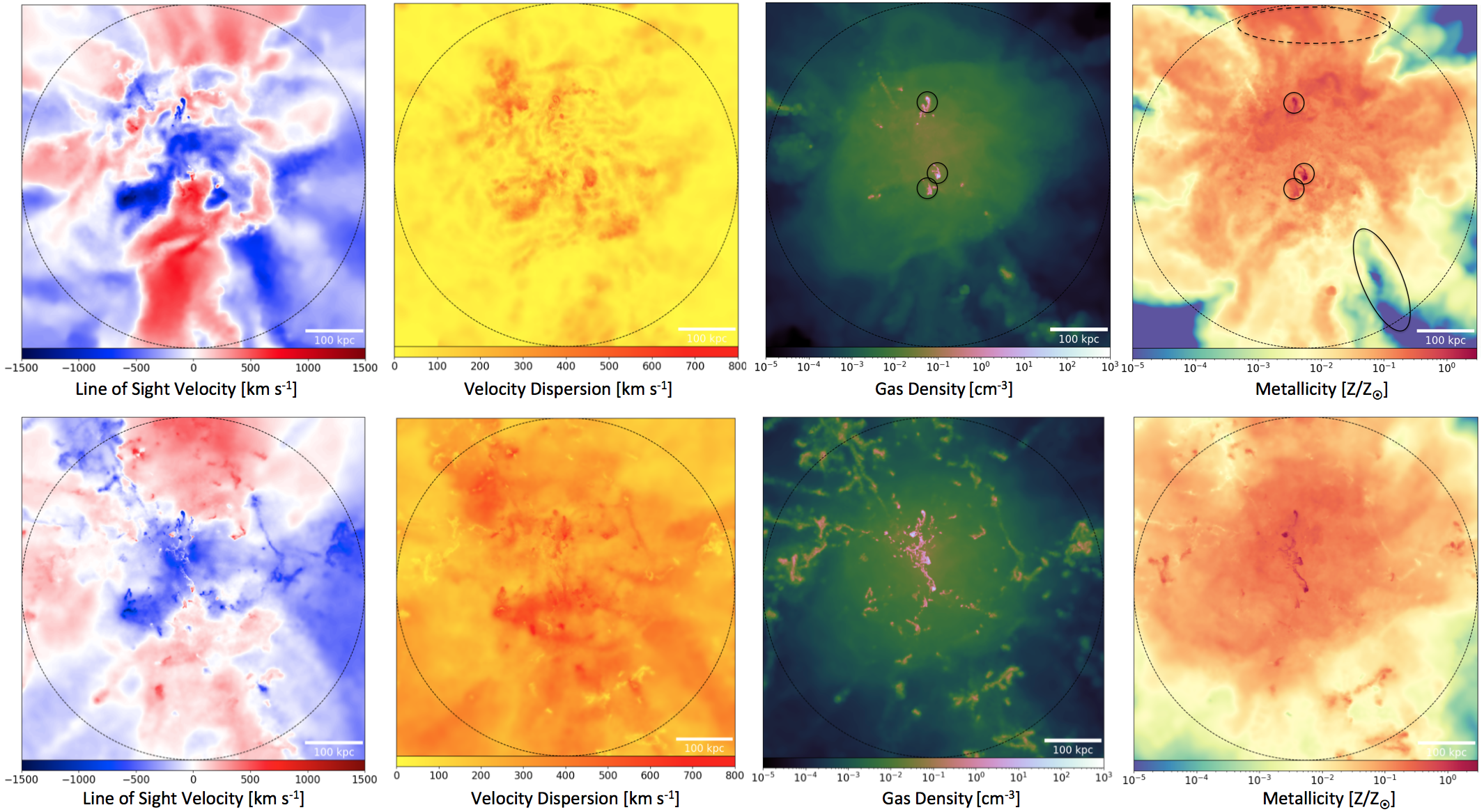}
\caption{Velocity field, velocity dispersion, gas density and metallicity in the CGM of one of the most massive
galaxies at $z = 3$ in the FABLE simulation. All quantities are mass-weighted.
The top panels show quantities averaged inside a slice that has a thickness of 5\% of the virial radius (on either side of the central galaxy), while in the bottom panels the quantities are averaged across a slice as thick as virial radius. The dashed circle shows the virial radius. The velocity and velocity dispersion scales are set to match those in Figure~\ref{fig:velocity}. In the top right panels, we mark examples of high metallicity and high density clumps with solid circles, the high metallicity diffuse outflow with a dashed black ellipse, and example of low metallicity inflow with a solid black ellipse.
\label{fig:fable}}
\end{figure*}

\subsection{Comparison with metallicities inferred from absorption systems}  \label{subsec:comp_abs}

It is also interesting to compare our findings on the metallicity of the CGM at z$\sim$3--4 with the results obtained
from absorption-line studies. While our results probe the inner regions of the CGM (because weighted by density and by proximity to the quasar's ionizing radiation), absorption lines usually trace outer regions \citep[e.g.][]{kurt05,joseph13,lau16}. For example, \citet{lau16} probe the CGM at a distance of 40 kpc and beyond. They find that the CGM is significantly enriched even beyond the estimated virial radius. Within the central 200 kpc, the median metallicity [M/H] is -0.6 dex ($\sim 0.25~Z_{\odot}$), which is only slightly lower than our results and it is expected because they span a larger radius. It is reassuring that emission and absorption line studies obtain consistent results for what concerns the metallicity of the CGM in the inner regions. However, \citet{lau16} also find that the ionization parameter increases with radius, which is the opposite trend with what inferred through the emission lines. The likely origin of the discrepancy
is that emission lines preferentially detect the densest clumps in the CGM (because of the squared dependence
on the density of the emission lines), while the absorption lines probe the bulk of the lower density gas in the CGM along
a specific line of sight.

\subsection{Comparison with Cosmological Simulations}  \label{subsec:simulations}

The metal enrichment of galactic haloes as well as the mixing with near-pristine inflows from the IGM is a complex phenomenon which is best investigated through cosmological simulations.
In the previous sections we have already discussed some of the results obtained by cosmological simulations \citep[e.g.][]{muratov17,hafen19}.

We have investigated the comparison with simulations more quantitatively by exploiting the FABLE simulation suite.
The FABLE (Feedback Acting on Baryons in Large-scale Environments) simulations \citep[][Bennett et al. in prep.]{henden18,henden19,henden19b} are a suite of state-of-the-art cosmological hydrodynamical simulations of galaxies, groups and clusters performed with the AREPO moving-mesh code \citep{springel10}. The simulations employ an updated set of physical models for AGN and supernovae feedback on top of the successful Illustris galaxy formation model \citep{genel14,vogelsberger14,sijacki15} to reproduce the present-day stellar and gas mass fractions of galaxy groups and clusters across a wide halo mass range.
For our comparisons we consider the most massive galaxies ($\mathrm{log(M_*/ M_{\odot})>11.0}$) of the FABLE cluster zoom-in simulations described in \citet{henden19} at z = 3, i.e. around the same redshift as the quasars in the MUSE sample. These galaxies are in dark matter halo masses in the range $\mathrm{13.03 <log(M_{200}/M_{\odot})<13.76}$, and have stellar masses in the range $\mathrm{11.01<log(M_*/ M_{\odot})<11.58}$.

Despite selecting the most massive galaxies and despite the size of the FABLE simulations, these galaxies are probably not as massive as the hosts of the very luminous quasars targeted by the MUSE surveys. Although we do not have the masses of the host galaxies of the quasars observed with MUSE, we can compare the black hole masses of the most massive galaxies selected in the FABLE with those measured in the MUSE quasar sample. The BH masses can be considered as a proxy of the galaxy masses, through the $\rm M_{BH}-\sigma$ relation. As illustrated in Figure.~\ref{fig:BH_mass}, the BHs in the MUSE quasar sample (blue histogram) are about an order of magnitude more massive than the quasars in the most massive galaxies selected in the FABLE simulations (orange histogram). Since the metallicity generally scales with mass, it is expected that the FABLE galaxies will be on average less metal rich than the observed sample.

As an example, Figure~\ref{fig:fable} shows the velocity, velocity dispersion, gas density and gas metallicity of the CGM in the halo of one of the most massive galaxies in the FABLE simulation at $z = 3$. All quantities are mass-weighted averages. The velocity and velocity dispersion scales are set to match those in Figure.~\ref{fig:velocity}.
The top panels show quantities averaged inside a slice that has a thickness of 5\% of the virial radius (on either side of the central galaxy), while in the bottom panels they are averaged across a slice as thick as virial radius.
These two cases should bracket what is seen in observations.
Indeed, the observations see the gas averaged in projection, however the velocity information allows us to isolate emitting gas clumps along the line-of-sight (although the stacking tends to average out all velocities). Moreover, regions (slices) closer to the quasar are illuminated by a stronger radiation field and therefore tend to emit more strongly and to weight more in the final averaging along the line-of-sight. Finally, the gas emissivity scales as the squared power of the gas density, hence (for a given radiation field) denser clumps dominate the emission of the nebular lines averaged along the line-of-sight. Taking into account all of these observational effects would imply a much more complex treatment of the simulations, which goes beyond the scope of this paper. Here we use the FABLE simulation simply to see whether the main observational features inferred from the MUSE data are generally consistent with the simulations, in a quantitative sense, but we leave a more detailed comparison to a future work.

The simulation clearly shows a complex distribution of both metal poor inflowing streams (e.g. the one traced with the black ellipse in Figure~\ref{fig:fable}), metal rich outflows (e.g. the region highlighted by the dashed black ellipse in Figure~\ref{fig:fable}), and more complex patterns. It is interesting to note that similar complex velocity patterns are seen in some of the Ly$\alpha$ velocity fields discussed above for individual objects. The velocity dispersion reaches high values, similar to those inferred from the observations.
In the central region, we see enhancement in velocity dispersion similar to Figure.~\ref{fig:velocity}, which may be a consequence of powerful nuclear outflows.

It is also interesting to note that the average gas density is obviously low across most of the CGM, but that there are several (metal rich) clumps around or even exceeding $\mathrm{10~cm^{-3}}$ (e.g. those marked with black solid circles in Figure~\ref{fig:fable}); these are likely dominating the emission that we see in the nebular lines, due to the quadratic dependence of the emissivity on the gas density, and therefore further justifying our choice of the density in the photoionization models. However, within the context of this paper, the main feature to note in the simulation is that there are several regions of the CGM within the virial radius, and even beyond it, that are enriched to metallicities approaching or even exceeding solar, as a result of prominent quasar outflows. At the same time very low metallicity CGM (less than 0.01 $Z_{\odot}$ and even down to $10^{-5}~Z_{\odot}$) is observed even inside half of the virial radius, close to the galaxy, as a consequence of denser streams from the IGM that manage to pierce into the hot halo.

\begin{figure*}[ht!]
\includegraphics[width=0.32\textwidth]{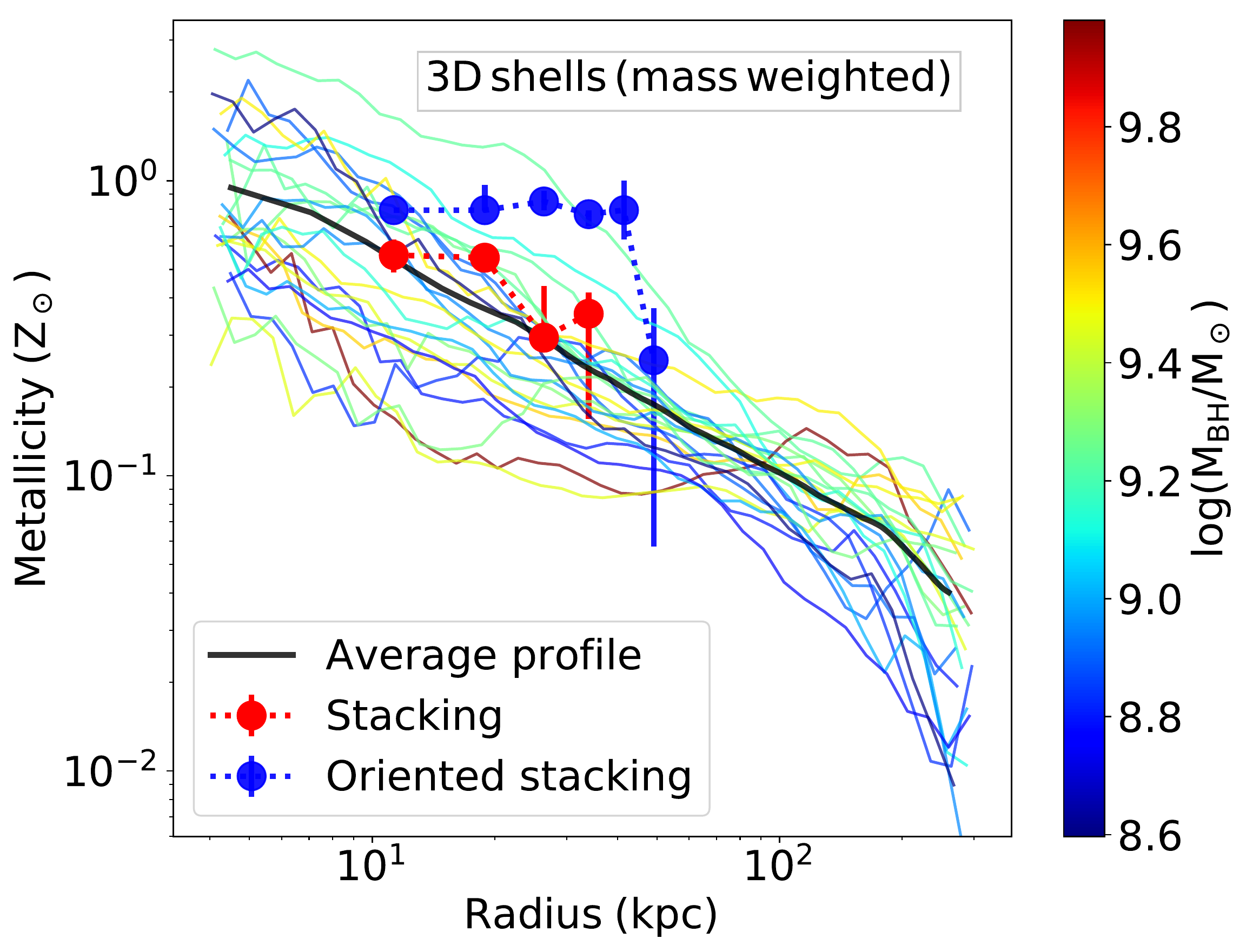}
\includegraphics[width=0.32\textwidth]{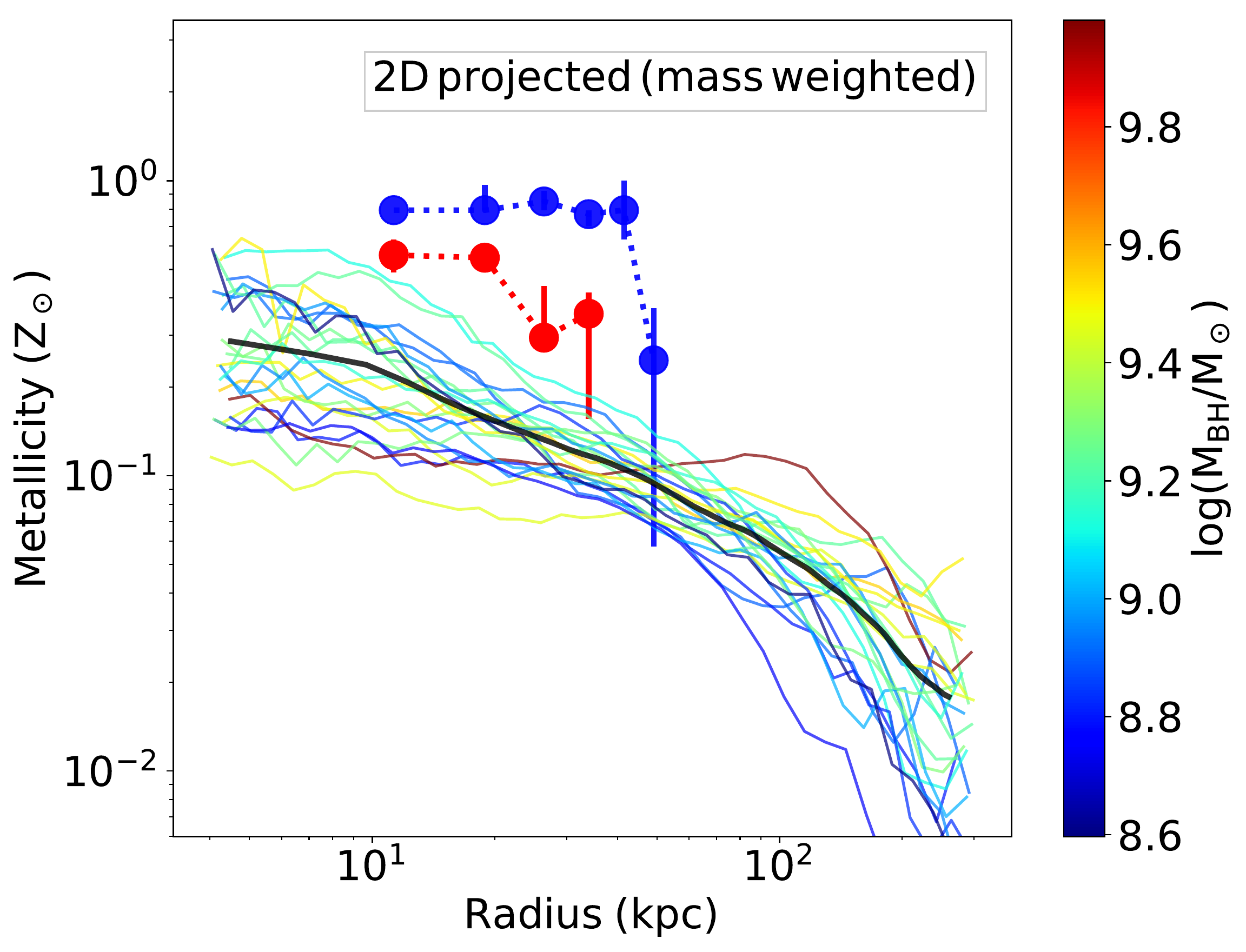}
\includegraphics[width=0.32\textwidth]{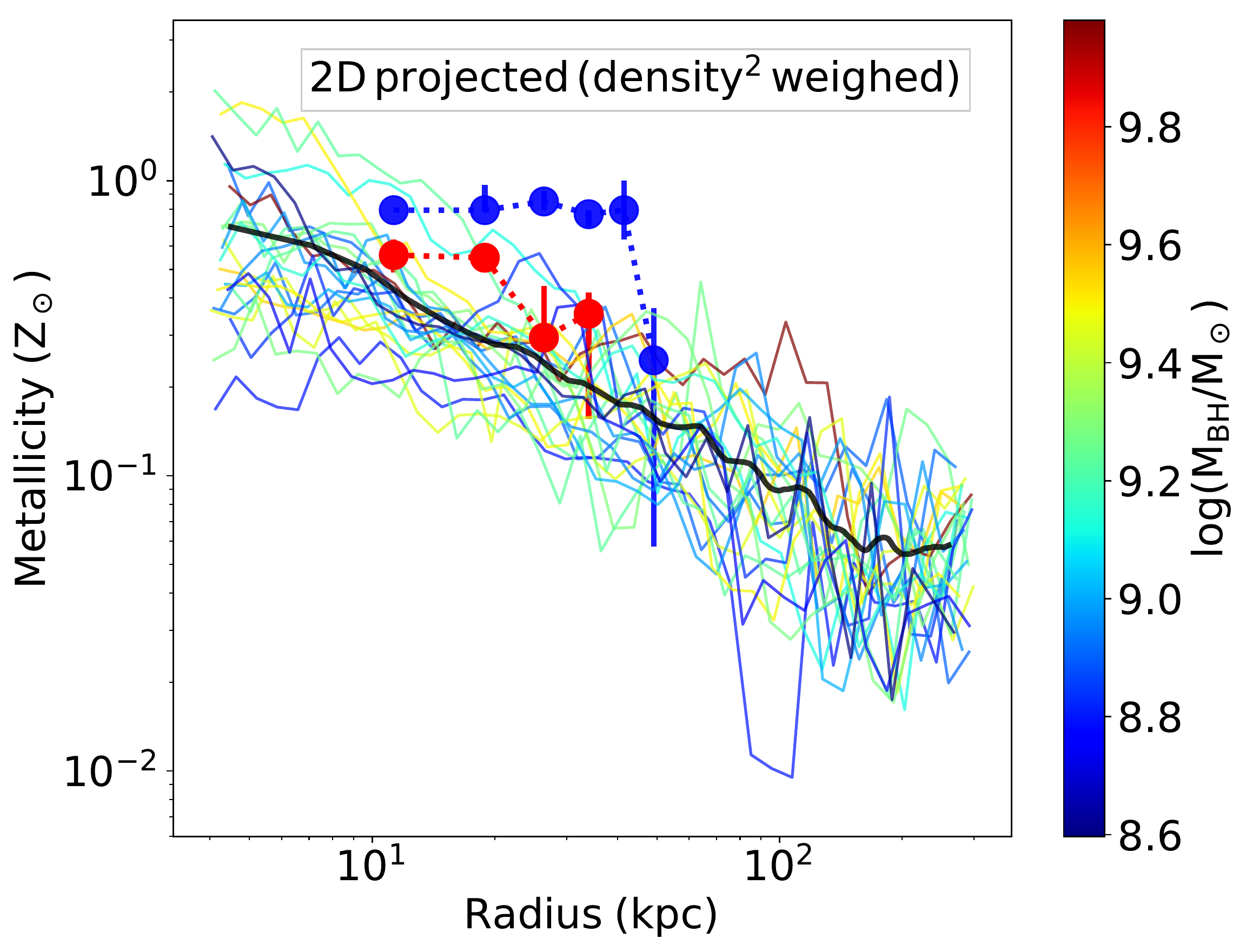}
\caption{Radial metallicity profiles of the CGM in galactic haloes at $z = 3$ from the FABLE cosmological simulations (lines color-coded by black hole mass) and radial metallicity inferred from the stacked MUSE cubes both for the global stacking (red symbols) and for the stacking of the cubes realigned along the extension of the Ly$\alpha$ extension (blue symbols). The average metallicity profile for all simulated haloes is shown by the black line.
In the leftmost panel the metallicity of the cosmological simulation is averaged
inside three-dimensional radial bins and mass-weighted. In the central panels the metallicity is averaged in
annuli projected along the line-of-sight and mass-weighted. In the rightmost panel the metallicity is averaged in
annuli projected along the line-of-sight and weighted by the square of the gas density (akin to the gas emissivity).
\label{fig:sim_vs_obs}}
\end{figure*}

Clearly, we do not have the sensitivity to map these complex metallicity structures in any of the individual observations in our sample.
However, we can compare the radial profiles obtained from our stacks with the radial metallicity profiles likewise extracted from the cosmological simulations. This comparison is illustrated in Figure~\ref{fig:sim_vs_obs}, where the lines show the metallicity profiles of the CGM in various galactic haloes at $z = 3$ in the simulation, color-coded by black hole mass, while the symbols show the metallicity inferred for our stacks at different radial points (red symbols for simple stacks, blue symbols for stacks obtained by re-orienting the cubes along the Ly$\alpha$ extension) obtained from the grids shown in Figures \ref{fig:stack1_metallicity}.
In the leftmost panel the metallicity of the cosmological simulation is averaged
inside three-dimensional radial bins and mass-weighted. In the central panel the metallicity is averaged in
annuli projected along the line-of-sight and mass-weighted. In the rightmost panel the metallicity is averaged in
annuli projected along the line-of-sight and weighted by the square of the gas density. The latter is expected to be the case closest to the observations, as the emissivity of these nebular lines scales quadratically with density (as long as the densities are lower than the critical densities, which is certainly the case for these transitions in the ISM and CGM), but still not exactly reproducing the case of the observations as this does not yet include the radial geometrical dilution of the quasar ionizing radiation. We have chosen to color-code the simulated haloes by black hole mass as this is supposed to provide a measure of the integrated AGN feedback, hence is expected to possibly scale with the CGM metal enrichment by quasar-driven outflows. However, the metallicity of the simulated massive haloes does not increases monotonically with BH mass;
there are haloes with very massive BH and low metallicity, as well as haloes with low mass BH and relatively high metalilcity. We do not find any obvious trend even when color-coding the haloes by halo mass (M$_{200}$, not shown).
The lack of clear correlation at this early epochs between metallicity and BH/halo mass reflects the fact that the halo CGM metallicity is strongly dependent on the recent history of accretion from the IGM and the recent enrichment (or lack thereof) by quasar-driven winds.
However, one should also take into account that here we have selected the 20 galaxies with the highest stellar mass, which corresponds to a fairly narrow dynamic range in halo mass.
A broader range of halo masses should be included to properly investigate the metallicity dependence on halo mass. However, it is interesting to note that the metallicities inferred from the stacked cubes are consistent with the upper envelope of the profiles obtained by the cosmological simulation. The `oriented stacking' profiles are characterized by higher metallicities (though still consistent with the most metal-rich halo in the simulation), not surpisingly, given that in this case we have probably aligned the haloes along the direction of quasar illumination and which is also probably the direction along which metal-loaded outflows are ejected. We also recall that the black holes of these most massive galaxies selected from the simulation are one order of magnitude less massive than those inferred for the quasars observed by MUSE, hence are expected to be less metal
enriched than in the observations. More massive simulated systems are expected to result in significantly
higher metallicities, even closer to those observed.

Note that the simulations do expect even super-solar metallicities for some haloes in their inner regions, suggesting that also the higher metallicity inferred  through the method of \cite{dors19} are plausible according to the cosmological simulations.

In the simulation the average radial metallicities drop below 0.1~$Z_\odot$ at a radial distance larger than about 100~kpc (about half of the virial radius  for these massive haloes). As discussed above, and as shown in Figure~\ref{fig:fable}, pockets of low metallicity gas are also likely present well within
the central regions of the halo. The presence of such low metallicity gas is observationally confirmed by the spaxels of Ly$\alpha$ emitting gas with no \civ, whose stacked spectrum does not show evidence of any metal line (Sect.~\ref{subsec:cgm_components} and Figure.~\ref{fig:lya_civ}).

Overall, given the complex physics involved in these physical processes, that are not simple to incorporate in the cosmological simulations, and given the uncertainties affecting the observations, especially for what concerns the nebular lines detections and metallicity measurements, and, most importantly, taking into account that the black holes (hence probably the haloes) in the simulations are about one order of magnitude less massive than in our sample
(Figure~\ref{fig:BH_mass}, hence the simulations are certainly probing haloes that are less enriched than our sample), there is good agreement between our observational results and the expectations from these zoom-in cosmological simulations.

Extended metal-line emission has been predicted by various other simulations \citep[e.g.][]{serena10b,bertone12,freeke13}.
It is beyond the scope of this paper to perform a detailed comparison with all simulations. Here however we note that the radial \civ\  profiles obtained by us are higher than obtained by \citet{freeke13}, possibly because they do not take into account the illumination by a quasar and only consider emission from intergalactic gas with densities $\mathrm{n_H\, < \, 0.1 \, cm^{-3}}$. Yet, \citet{freeke13} also show that the flux-weighted metallicity may be biased high compared to the mass-weighted metallicity because metal-line emission is proportional to the local metallicity.

\subsection{What Are the Mechanisms Powering the UV Emission Line Nebulae?}

\begin{figure}[ht!]
\plotone{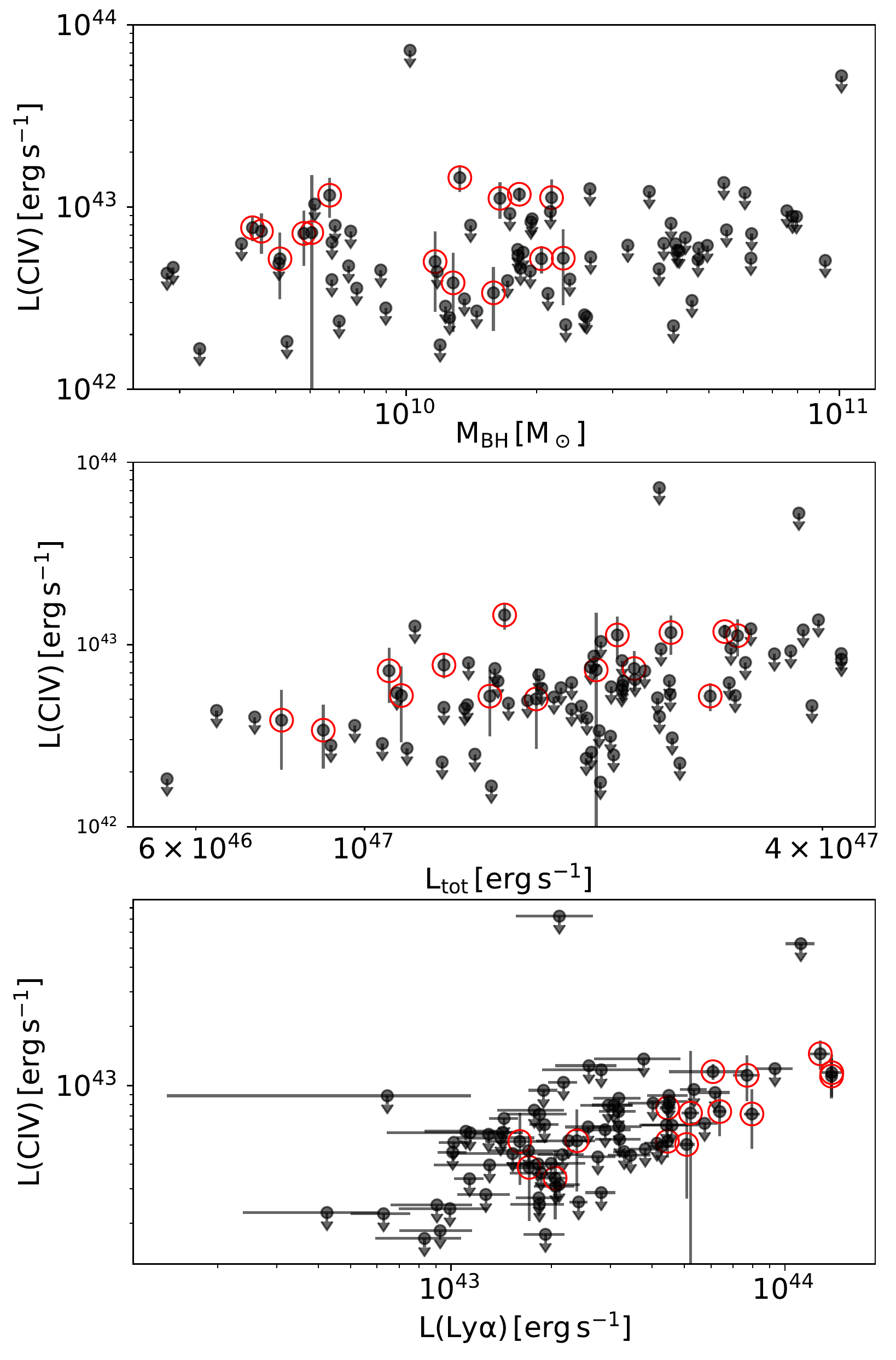}
\caption{Upper panel: the central black hole mass vs. \civ\ luminosity. Middle panel: total black hole luminosity vs. \civ\ luminosity. Bottom panel: the relation between Ly$\alpha$ luminosity and \civ\ luminosity. In all three panels, the red circles denote the detections of \civ. All other black dots are upper limits.  \label{fig:params_civ}}
\end{figure}

Figure~\ref{fig:params_civ} shows the luminosity of \civ\ versus black hole mass (upper panel), total luminosity of quasars (middle panel) and Ly$\alpha$ luminosity (bottom panel). The total luminosity of quasars is calculated based on a correction factor to the luminosity at 3000\AA\ \citep{trakhtenbrot12}. In all three panels, the objects with the detection of \civ\ are denoted by red circles. Other black symbols provide upper limits.
We see no correlation between \civ\ luminosity and black hole mass.
There is possibly a mild correlation between \civ\ luminosity and total luminosity, but the scatter is large.
This result is not surprising considering the objects in our sample are all at the high mass and high luminosity end of the quasar population (hence limited dynamic range) and also a consequence of the flickering nature of AGNs. However, despite the large number of upper limits and only a limited number of detections, we observe  a correlation between \civ\ luminosity and Ly$\alpha$ luminosity, as shown in the bottom panel of Figure~\ref{fig:params_civ}.
This suggests, as expected, that the Ly$\alpha$ and \civ\ emissions result from photoionization from the same sources, i.e. the quasar radiation field, which is indeed the expectation based on which these very luminous quasars were selected in order to enhance the probability of detecting Ly$\alpha$ haloes. The correlation between CIV and Ly$\alpha$ is maintained despite the quasar variability because of the much larger timescales associated with the photoionization of the CGM, involving the light travel time to several tens of kpc in the CGM, which smooths away the AGN variability \citep[e.g.][]{gilli00}.

\section{Summary and conclusions}
\label{sec:conclusions}
Recent MUSE observations have detected extended Ly$\alpha$ nebulae around quasars at redshift 3$-$6.
In this paper we have combined the two samples of quasars observed with MUSE presented in \citet{borisova16} and \citet{fabrizio19}, the vast majority of which show extended Ly$\alpha$ nebulae, with the goal of detecting diffuse \civ, \heii, and \ciii\ emission to constrain the physical and chemical properties of the CGM. Our main findings are summarized as follows:

\begin{itemize}

\item Extended \civ\ nebulae are found in the haloes of 15 individual quasars ($\sim 19\%$ of the whole sample, most of which radio quiet). Extended \heii\ is detected in 10 objects ($\sim 13\%$). Extended \ciii\ emission is much fainter and detected only in 4 objects ($5\%$).

\item Morphologically, the \civ, \heii, and \ciii\ nebulae are more compact than the Ly$\alpha$ nebulae, and are characterized by a range of diverse shapes. The compactness of these nebulae is mostly due to the weakness of these lines, but is likely also associated with a radial drop in ionization conditions and metallicity (as inferred from the stacking).

\item The kinematics of the \civ \ and \heii \ nebulae (no kinematic information could be inferred for the \ciii \ nebulae) are similar to their parent Ly$\alpha$ nebulae. The central region is characterized by large velocity dispersion, pointing at quasar-driven outflows, but in the outer regions \civ\  and \heii \ appear to trace the dynamically more quiescent CGM.

\item We obtain average maps of Ly$\alpha$, \civ, \heii, and \ciii\ by stacking the cubes of all quasars together (after re-aligning them in redshift and spatially). We also stack the cubes after re-aligning them along the direction of the primary Ly$\alpha$ extension.
We detect these lines out to a radial distance of about 35--42 kpc from the (stacked) quasar.
\civ \ and \heii \ show a steeper decline with radius than Ly$\alpha$. The stacked \ciii \ radial profile is shallower than \civ \ and \heii \ and similar to the radial profile of Ly$\alpha$.
These trends suggest that the decline of \civ \ and \heii \ is mostly due to a decline in ionization parameter, while the average metallicity changes little with radius at these distances.

\item We use AGN photoionization models to constrain the CGM properties through the observed line ratios. We find that the ionization parameter declines rapidly with radial distance, while the metallicity tend to vary more slowly and is between 0.5~$Z_{\odot}$ and $1~Z_{\odot}$ out to $\sim 40$~kpc.

\item The inferred metallicities are somewhat model dependent. We have shown that other recent photoionization models \citep[e.g. ][]{dors19} would give even higher (super-solar) metallicities.

\item We have also stacked the extended emission spectra by splitting them based on their black hole masses. We have found that haloes hosting more massive black holes are characterized by slightly higher metallicities than haloes hosting less massive black holes. This is not an unexpected trend, given that more massive black holes are hosted in more massive galaxies, which are therefore more metal-rich (based on the mass-metallicity relation), whose outflows are presumably also more metal rich.

\item By stacking all spaxels in the stack that show no evidence for \civ\ we obtain a high signal-to-noise stacked spectrum that still fails to show any hint of metal lines. Therefore, these regions are very metal poor.

\item By combining these results we conclude that the Ly$\alpha$ nebulae observed around quasars are characterized by two components: a component of the CGM that has been highly enriched by galactic outflows and a very low metallicity component associated with streams accreting from the IGM.

\item We have compared our observational results with the metallicity maps of the CGM of haloes associated with massive galaxies at z = 3 in the FABLE cosmological simulations. Such simulated haloes appear indeed characterized by very high metallicities (close to solar or even super-solar) with irregular distribution, but primarily associated with galactic outflows, even out to and beyond the viral radius (i.e. $\sim$200 kpc), but also by low metallicity streams entering into the halo down to less than half a virial radius. The average radial profiles of the metallicity from the simulated massive haloes show a good agreement with our observational results. Overall, these results suggest that the observations of Ly$\alpha$ haloes surrounding quasars are capturing the properties of the CGM, associated with metal enrichment by galactic outflows and near-pristine accretion from the IGM, that are expected by the cosmological simulations, even at a quantitative level.

\item Finally, we investigate the correlation of the \civ\ luminosity with the quasar properties.
We find a positive correlation between \civ\ luminosity and Ly$\alpha$ luminosity. \civ\ and Ly$\alpha$ nebulae are likely powered by the same central engine, i.e. the quasar ionizing radiation. However, due to the flickering and rapidly varying properties of quasars, they are not correlated directly with other properties of central quasars.
\end{itemize}

\begin{longrotatetable}
\begin{deluxetable*}{ccccccccccc}
\tablecaption{Quasar Sample and Properties\label{tab:sample}}
\tabletypesize{\scriptsize}
\tablehead{
\colhead{ID} & \colhead{Quasar} &
\colhead{RA} & \colhead{Dec} &
\colhead{Seeing} & \colhead{$z_{Ly\alpha}$ \tablenotemark{a}} &
\colhead{Exp. T.} & \colhead{$\mathrm{L_{Ly\alpha}}$ \tablenotemark{b}} &
\colhead{$\mathrm{M_{BH}}$ \tablenotemark{c}} & \colhead{$\mathrm{L_{bol}}$ \tablenotemark{d}} & \colhead{Class \tablenotemark{e}} \\
\colhead{} & \colhead{} & \colhead{(J2000)} & \colhead{(J2000)} &
\colhead{(arcsec)} & \colhead{} & \colhead{(min)} & \colhead{($\mathrm{10^{43}\, erg\, s^{-1}}$)} &
\colhead{($\mathrm{10^{10}\, M_\odot}$)} & \colhead{($\mathrm{10^{47}\, erg\, s^{-1}}$)} & \colhead{}
}
\startdata
1 & SDSS J2319 1040 & 23:19:34.800 & -10:40:36.00 & 1.4 & 3.171 & 42.44 & 1.53 & 0.87 & 1.27 & RQ \\
2 & UM 24 & 00:15:27.400 & +06:40:12.00 & 1.62 & 3.165 & 42.29 & 2.82 & 6.05 & 3.77 & RQ \\
3 & J 0525 233 & 05:25:06.500 & -23:38:10.00 & 0.86 & 3.119 & 42.88 & 1.29 & 1.86 & 2.18 & RL \\
4 & Q 0347 383 & 03:49:43.700 & -38:10:31.00 & 1.4 & 3.231 & 84.71 & 2.06 & 4.57 & 2.54 & RQ \\
5 & SDSS J0817 1053 & 08:17:52.099 & +10:53:29.68 & 1.47 & 3.332 & 42.44 & 3.18 & 3.25 & 1.87 & RQ \\
6 & SDSS J0947 1421 & 09:47:34.200 & +14:21:17.00 & 1.21 & 3.069 & 42.27 & 0.64 & 7.96 & 3.46 & RQ \\
7 & SDSS J1209 1138 & 12:09:18.000 & +11:38:31.00 & 1.22 & 3.118 & 42.23 & 1.89 & 2.15 & 2.45 & RQ \\
8 & UM 683 & 03:36:26.900 & -20:19:39.00 & 1.05 & 3.132 & 85.02 & 4.45 & 0.44 & 1.27 & RQ \\
9 & Q 0956 1217 & 09:58:52.200 & +12:02:45.00 & 1.12 & 3.311 & 42.25 & 9.33 & 3.64 & 3.22 & RQ \\
10 & SDSS J1025 0452 & 10:25:09.600 & +04:52:46.00 & 1.1 & 3.242 & 42.58 & 7.7 & 2.17 & 2.15 & RQ \\
11 & Q N1097 1 & 02:46:34.200 & -30:04:55.00 & 1.18 & 3.099 & 42.47 & 1.71 & 1.28 & 0.78 & RQ \\
12 & SDSS J1019 0254 & 10:19:08.255 & +02:54:31.94 & 1.22 & 3.394 & 42.1 & 4.03 & 4.09 & 2.18 & RQ \\
13 & PKS 1017 109 & 10:20:10.000 & +10:40:02.00 & 0.91 & 3.167 & 242.13 & 4.45 & 2.05 & 2.85 & RQ \\
14 & SDSS J2100 0641 & 21:00:25.030 & -06:41:45.00 & 0.85 & 3.133 & 42.49 & 1.11 & 1.81 & 2.11 & RQ \\
15 & SDSS J1550 0537 & 15:50:36.806 & +05:37:50.07 & 1.01 & 3.145 & 42.33 & 5.75 & 0.67 & 2.27 & RQ \\
16 & SDSS J2348 1041 & 23:48:56.488 & -10:41:31.17 & 1.72 & 3.185 & 42.08 & 1.44 & 4.41 & 1.69 & RQ \\
17 & SDSS J0001 0956 & 00:01:44.886 & -09:56:30.83 & 1.81 & 3.348 & 42.24 & 1.84 & 2.61 & 1.4 & RQ \\
18 & SDSS J1557 1540 & 15:57:43.300 & +15:40:20.00 & 0.81 & 3.291 & 42.13 & 12.76 & 1.33 & 1.53 & RQ \\
19 & SDSS J1307 1230 & 13:07:10.200 & +12:30:21.00 & 0.89 & 3.225 & 42.25 & 1.84 & 6.27 & 2.34 & RQ \\
20 & SDSS J1429 0145 & 14:29:03.033 & -01:45:19.00 & 0.79 & 3.424 & 42.4 & 2.57 & 4.95 & 3.02 & RQ \\
21 & CT 669 & 20:34:26.300 & -35:37:27.00 & 0.74 & 3.219 & 42.28 & 6.38 & 0.46 & 2.26 & RQ \\
22 & Q 2139 4434 & 21:42:25.900 & -44:20:18.00 & 1.02 & 3.229 & 42.32 & 5.34 & 7.57 & 3.03 & -- \\
23 & Q 2138 4427 & 21:41:59.500 & -44:13:26.00 & 0.81 & 3.139 & 42.12 & 4.14 & 9.28 & 2.43 & -- \\
24 & SDSS J1342 1702 & 13:42:33.200 & +17:02:46.00 & 1.03 & 3.058 & 43.03 & 4.41 & 0.42 & 1.5 & RQ \\
25 & SDSS J1337 0218 & 13:37:57.900 & +02:18:21.00 & 0.75 & 3.343 & 42.39 & 1.02 & 4.72 & 1.77 & RQ \\
26 & Q 2204 408 & 22:07:34.300 & -40:36:57.00 & 0.84 & 3.185 & 42.35 & 11.16 & 10.11 & 3.72 & -- \\
27 & Q 2348 4025 & 23:51:16.100 & -40:08:36.00 & 0.87 & 3.331 & 42.7 & 3.2 & 2.67 & 2.52 & -- \\
28 & Q 0042 269 & 00:44:52.300 & -26:40:09.00 & 0.8 & 3.36 & 42.49 & 1.41 & 1.81 & 1.1 & RQ \\
29 & Q 0115 30 & 01:17:34.000 & -29:46:29.00 & 0.91 & 3.227 & 42.42 & 1.14 & 4.27 & 1.71 & RQ \\
30 & SDSS J1427 0029 & 14:27:55.800 & -00:29:51.00 & 0.95 & 3.359 & 42.36 & 2.17 & 0.61 & 2.04 & RQ \\
31 & UM 670 & 01:17:23.300 & -08:41:32.00 & 0.8 & 3.204 & 42.63 & 2.25 & 6.25 & 3.07 & RQ \\
32 & Q 0058 292 & 01:01:04.700 & -28:58:03.00 & 1.42 & 3.1 & 42.25 & 2.59 & 2.66 & 1.16 & RQ \\
33 & Q 0140 306 & 01:42:54.700 & -30:23:45.00 & 1.39 & 3.132 & 42.21 & 1.61 & 0.51 & 1.46 & RL \\
34 & Q 0057 3948 & 00:59:53.200 & -39:31:58.00 & 1.09 & 3.251 & 42.27 & 1.27 & 0.9 & 0.9 & RQ \\
35 & CTS C22 31 & 02:04:35.500 & -45:59:23.00 & 1.18 & 3.247 & 42.12 & 2.75 & 0.28 & 0.64 & -- \\
36 & Q 0052 3901A & 00:54:45.400 & -38:44:15.00 & 1.25 & 3.202 & 42.17 & 1.71 & 0.29 & 1.37 & RL \\
37 & UM672 & 01:34:38.600 & -19:32:06.00 & 1.87 & 3.128 & 81.64 & 0.83 & 0.33 & 1.47 & RL \\
38 & SDSS J0125 1027 & 01:25:30.900 & -10:27:39.00 & 0.74 & 3.351 & 42.28 & 1.14 & 2.12 & 2.03 & RQ \\
39 & SDSS J0100 2105 & 01:00:27.661 & +21:05:41.57 & 0.73 & 3.096 & 42.08 & 4.27 & 0.51 & 1.64 & RQ \\
40 & SDSS J0250 0757 & 02:50:21.800 & -07:57:50.00 & 0.67 & 3.338 & 42.46 & 1.99 & 2.39 & 2.44 & RQ \\
41 & SDSS J0154 0730 & 01:54:40.328 & -07:30:31.85 & 0.75 & 3.335 & 42.45 & 2.16 & 1.93 & 1.35 & RQ \\
42 & SDSS J0219 0215 & 02:19:38.732 & -02:15:40.47 & 0.64 & 3.034 & 42.42 & 3.82 & 0.74 & 1.55 & RQ \\
43 & CTSH22 05 & 01:48:18.130 & -53:27:02.00 & 1.34 & 3.128 & 41.96 & 3.17 & 0.74 & 1.48 & -- \\
44 & SDSS J2321 1558 & 23:21:54.980 & +15:58:34.24 & 1.61 & 3.236 & 42.14 & 1.43 & 4.23 & 1.81 & RQ \\
45 & FBQS J2334 0908 & 23:34:46.400 & -09:08:12.24 & 1.18 & 3.358 & 256.69 & 0.63 & 4.14 & 2.6 & RL \\
46 & Q2355 0108 & 23:58:08.540 & +01:25:07.20 & 1.23 & 3.398 & 42.19 & 1.91 & 3.93 & 2.52 & RQ \\
47 & 6dF J0032 0414 & 00:32:05.380 & -04:14:16.21 & 1.59 & 3.162 & 42.18 & 7.96 & 0.58 & 1.08 & RL \\
48 & UM 679 & 02:51:48.060 & -18:14:29.00 & 0.96 & 3.219 & 42.21 & 1.83 & 1.46 & 1.14 & RQ \\
49 & PKS0537 286 & 05:39:54.267 & -28:39:56.00 & 0.74 & 3.138 & 42.4 & 2.05 & 1.59 & 0.88 & RL \\
50 & SDSS J0819 0823 & 08:19:40.580 & +08:23:57.98 & 0.73 & 3.205 & 42.01 & 13.79 & 1.65 & 3.09 & RQ \\
51 & SDSS J0814 1950 & 08:14:53.449 & +19:50:18.62 & 0.75 & 3.136 & 42.1 & 2.38 & 2.3 & 1.12 & RL \\
52 & SDSS J0827 0300 & 08:27:21.968 & +03:00:54.74 & 0.96 & 3.137 & 41.96 & 0.91 & 1.26 & 2.13 & RL \\
53 & SDSS J0905 0410 & 09:05:49.058 & +04:10:10.15 & 0.93 & 3.164 & 42.1 & 0.93 & 0.53 & 0.55 & RL \\
54 & S31013 20 & 10:16:44.319 & +20:37:47.29 & 0.92 & 3.111 & 42.04 & 1.82 & 0.67 & 0.72 & RL \\
55 & SDSS J1032 1206 & 10:32:12.886 & +12:06:12.83 & 0.89 & 3.191 & 42.15 & 2.96 & 0.69 & 1.37 & RL \\
56 & TEX1033 137 & 10:36:26.886 & +13:26:51.75 & 0.7 & 3.095 & 41.98 & 5.21 & 0.61 & 2.02 & RL \\
57 & SDSS J1057 0139 & 10:57:13.250 & -01:39:13.79 & 1.31 & 3.454 & 42.02 & 1.77 & 5.49 & 1.98 & RL \\
58 & Q1205 30 & 12:08:12.730 & -30:31:07.00 & 0.63 & 3.048 & 42.24 & 2.11 & 1.36 & 2.11 & RQ \\
59 & LBQS1244 1129 & 12:46:40.370 & +11:13:02.92 & 1.45 & 3.156 & 42.03 & 2.89 & 4.73 & 2.18 & RQ \\
60 & SDSS J1243 0720 & 12:43:53.960 & +07:20:15.47 & 1.29 & 3.179 & 42.01 & 1.86 & 0.77 & 0.97 & RL \\
61 & LBQS1209 1524 & 12:12:32.040 & +15:07:25.63 & 1.42 & 3.066 & 42.08 & 2.11 & 1.02 & 2.44 & RQ \\
62 & CTS G18 01 & 00:41:31.4 & -49:36:11.9 & 1.08 & 3.249 & 48.32 & 3.78 & 5.41 & 3.95 & RQ \\
63 & Q0041 2638 & 00:43:42.7 & -26:22:10.9 & 1.14 & 3.076 & 54.09 & 0.42 & 2.33 & 1.26 & RQ \\
64 & Q0042 2627 & 00:44:33.5 & -26:11:25.9 & 1.18 & 3.304 & 48.85 & 2.82 & 1.23 & 1.06 & RQ \\
65 & Q0055 269 & 00:57:58.1 & -26:43:15.8 & 1.02 & 3.659 & 551.25 & 6.09 & 1.82 & 2.98 & RQ \\
66 & UM669 & 01:05:16.7 & -18:46:41.9 & 1.31 & 3.038 & 54.1 & 3.45 & 1.18 & 1.87 & RQ \\
67 & J0124 0044 & 01:24:04.0 & 00:44:33.5 & 0.82 & 3.841 & 104.1 & 4.59 & 4.19 & 2.19 & RQ \\
68 & UM678 & 02:51:40.4 & -22:00:28.3 & 0.72 & 3.208 & 54.56 & 0.99 & 0.7 & 1.96 & RQ \\
69 & CTS B27 07 & 04:45:33.1 & -40:48:42.8 & 0.59 & 3.152 & 49.96 & 1.92 & 1.2 & 2.04 & RQ \\
70 & CTS A31 05 & 05:17:42.1 & -37:54:45.9 & 0.72 & 3.046 & 54.72 & 3.18 & 1.95 & 2.0 & RQ \\
71 & CT 656 & 06:00:08.7 & -50:40:30.1 & 0.7 & 3.154 & 54.37 & 1.01 & 3.84 & 1.93 & RQ \\
72 & AWL 11 & 06:43:26.9 & -50:41:12.9 & 0.63 & 3.118 & 69.04 & 2.41 & 2.58 & 1.99 & RQ \\
73 & HE0940 1050 & 09:42:53.6 & -11:04:26.0 & 0.74 & 3.091 & 48.93 & 4.49 & 7.8 & 4.23 & RQ \\
74 & BRI1108 07 & 11:11:13.7 & -08:04:03.0 & 0.98 & 3.935 & 104.29 & 5.08 & 1.17 & 1.68 & RQ \\
75 & CTS R07 04 & 11:13:50.1 & -15:33:40.2 & 0.94 & 3.366 & 48.7 & 13.82 & 0.67 & 2.53 & RQ \\
76 & Q1317 0507 & 13:20:29.8 & -05:23:34.2 & 0.94 & 3.719 & 421.51 & 6.18 & 1.74 & 3.63 & RQ \\
77 & Q1621 0042 & 16:21:16.7 & -00:42:48.2 & 0.85 & 3.707 & 198.46 & 3.07 & 1.41 & 3.17 & RQ \\
78 & CTS A11 09 & 22:53:10.7 & -36:58:15.9 & 0.76 & 3.147 & 54.49 & 1.3 & 1.71 & 1.96 & RQ \\
79 & PKS1937 101 & 19:39:57.4 & -10:02:39.9 & 0.75 & 3.791 & 155.68 & 4.51 & 1.94 & 4.24 & RL \\
80 & QB2000 330 & 20:03:24.1 & -32:51:45.9 & 0.96 & 3.789 & 528.28 & 3.3 & 1.84 & 3.88 & RL \\
\enddata
\tablenotetext{a}{$Ly\alpha$ redshift measured from diffuse $Ly\alpha$ nabulae.}
\tablenotetext{b}{$Ly\alpha$ luminosity measured from the PSF subtracted datacubes with circular aperture photometry with a radius of $2''$.}
\tablenotetext{c}{Black hole mass computed from quasar continuum luminosity at 1450\AA\ and FWHM of \civ, following the method of \citet{trakhtenbrot12}.}
\tablenotetext{d}{Quasar bolometric luminosity corrected by continuum luminosity at 3000\AA\ based on the method of \citet{trakhtenbrot12}. }
\tablenotetext{e}{RQ: radio quiet quasar; RL: radio loud quasar; the 6 sources labeled by -- are not covered by any radio survey.}
\end{deluxetable*}
\end{longrotatetable}


\acknowledgments
We acknowledge support from the National Science Foundation of China (11721303, 11890693) and the National Key R\&D Program of China (2016YFA0400703). RM acknowledges ERC Advanced Grant 695671 ``QUENCH" and support by the Science and Technology Facilities Council (STFC). JB and DS acknowledge support from the Science, Technology and Facilities Council (STFC) and the ERC Starting Grant 638707 ``Black holes and their
host galaxies: co-evolution across cosmic time”.


\appendix
\counterwithin{figure}{section}

\section{Pseudo-NB images for all objects}
\label{appen:pseudoNB}
\begin{figure*}[p]
\plotone{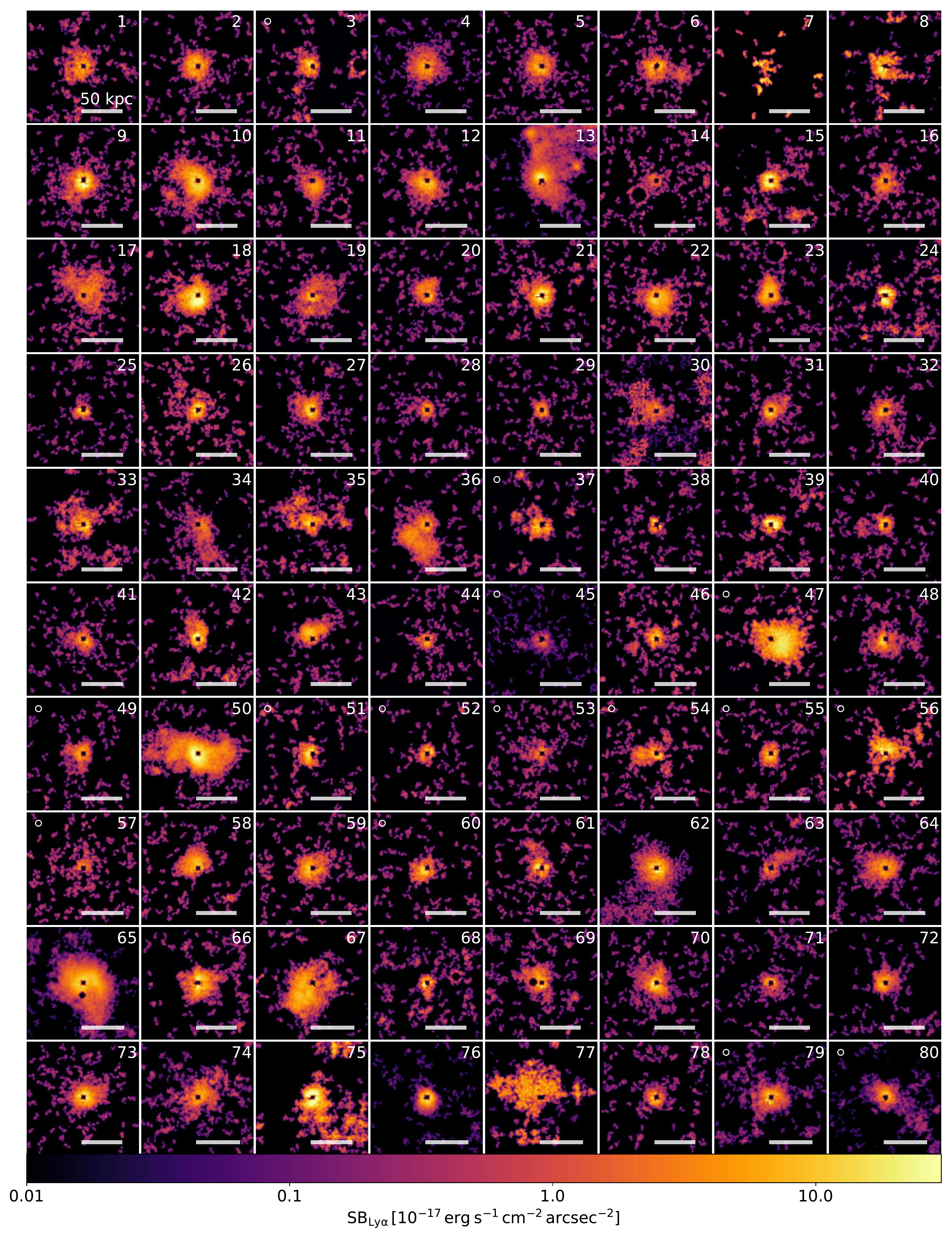}
\caption{Pseudo-NB images of the Ly$\alpha$ emission around quasars. In each panel, the original position of each quasar is masked by a small ($1''\times1''$) black square. Each thumbnail has a size of $20''\times20''$, corresponding to about 150 kpc at the median redshift of the sample.
\label{fig:lya_halos}}
\end{figure*}

In this section, we show the pseudo-NB images for all objects. In Figure~\ref{fig:lya_halos}, we show the atlas of the detected Ly$\alpha$ nebulae for the 80 quasars. These systems have very different shapes and sizes. Several of the Ly$\alpha$ nebulae are symmetric with physical scales smaller than 50 kpc. Several very extended objects (e.g., No. 13, 47, 50, 65, etc.) show very asymmetric emission. We compare these nebulae with those from \citet{borisova16} and \citet{fabrizio19}, and find that they broadly agree with each other.

\begin{figure*}[ht!]
\plotone{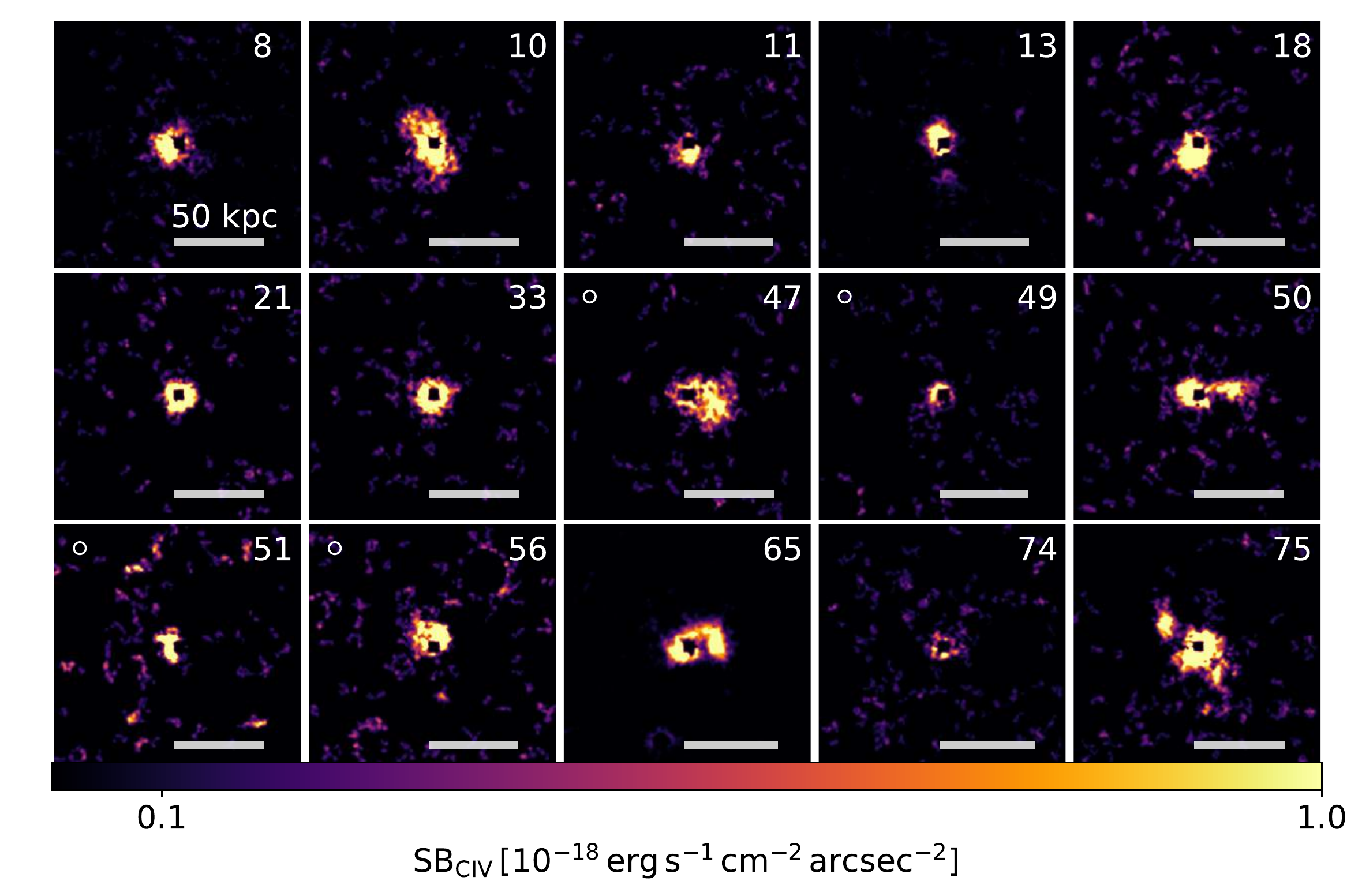}
\caption{The same as Figure~\ref{fig:lya_halos}, but for \civ. \label{fig:civ_halos}}
\end{figure*}

\begin{figure*}[ht!]
\plotone{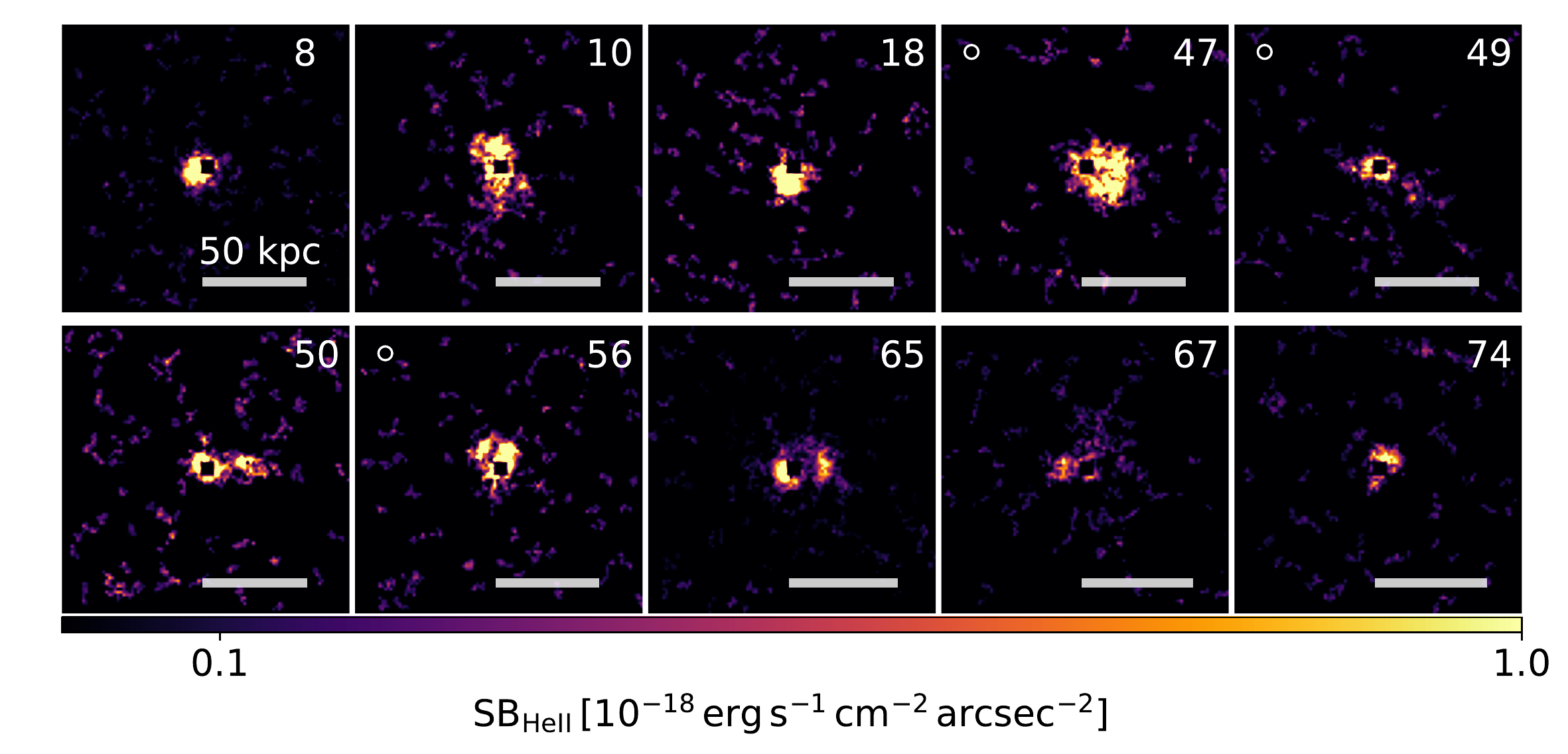}
\caption{The same as Figure~\ref{fig:lya_halos}, but for \heii. \label{fig:heii_halos}}
\end{figure*}

\begin{figure*}[ht!]
\plotone{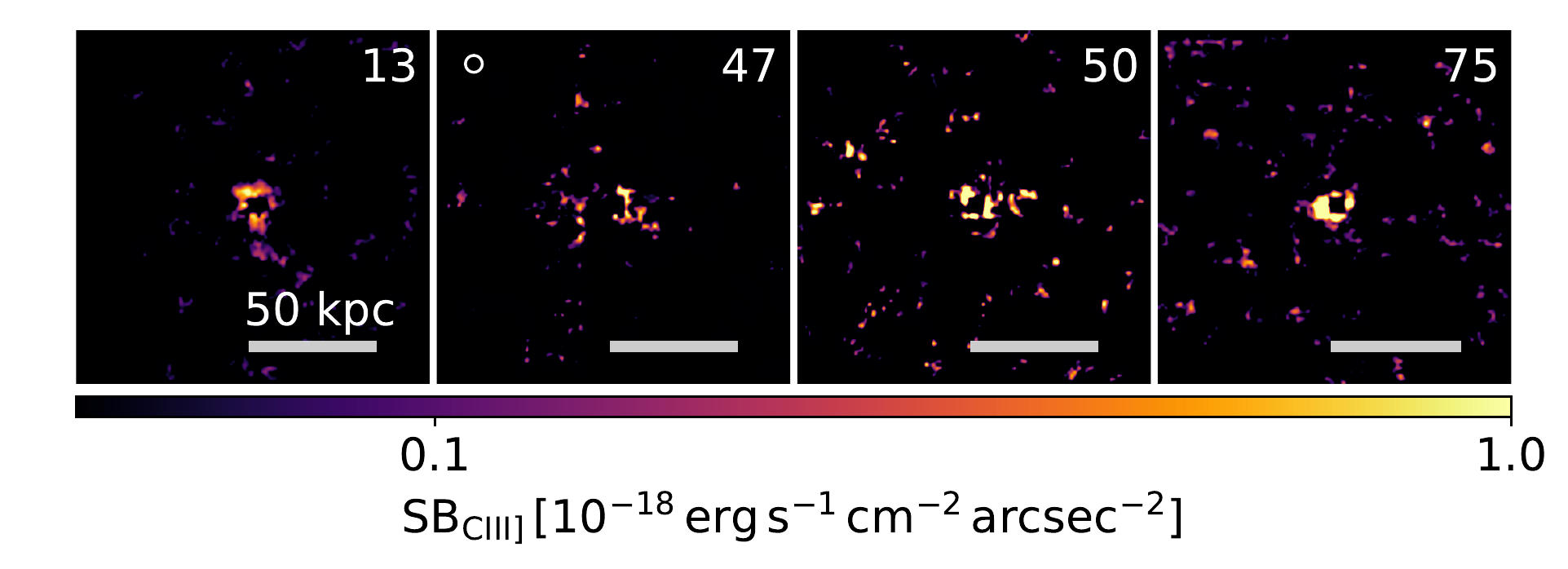}
\caption{The same as Figure~\ref{fig:lya_halos}, but for \ciii. \label{fig:ciii_halos}}
\end{figure*}

Out of 80 quasars, 15 are detected with extended \civ\ emission, as shown in Figure~\ref{fig:civ_halos}. These \civ\ nebulae are more compact than their Ly$\alpha$ counterparts. Ten objects are detected with extended \heii\ emission, as shown in Figure~\ref{fig:heii_halos}. Three objects are radio loud, and the others are radio quiet. The \heii\ morphology is generally similar to that of \civ. Only 4 objects are detected with extended \ciii\ emission, with 1 radio loud object, as shown in Figure~\ref{fig:ciii_halos}. For objects at redshift 3$\sim$4, the \ciii\ line highly contaminated by OH sky lines.

\bibliographystyle{aasjournal}
\bibliography{ref}

\begin{thebibliography}{}
\expandafter\ifx\csname natexlab\endcsname\relax\def\natexlab#1{#1}\fi
\providecommand{\url}[1]{\href{#1}{#1}}

\bibitem[{{Adelberger} {et~al.}(2005){Adelberger}, {Shapley}, {Steidel},
  {Pettini}, {Erb}, \& {Reddy}}]{kurt05}
{Adelberger}, K.~L., {Shapley}, A.~E., {Steidel}, C.~C., {et~al.} 2005, \apj,
  629, 636

\bibitem[{{Anderson} {et~al.}(2016){Anderson}, {Churazov}, \&
  {Bregman}}]{anderson16}
{Anderson}, M.~E., {Churazov}, E., \& {Bregman}, J.~N. 2016, \mnras, 455, 227

\bibitem[{{Arrigoni Battaia} {et~al.}(2019{\natexlab{a}}){Arrigoni Battaia},
  {Hennawi}, {Prochaska}, {O{\~n}orbe}, {Farina}, {Cantalupo}, \&
  {Lusso}}]{fabrizio19}
{Arrigoni Battaia}, F., {Hennawi}, J.~F., {Prochaska}, J.~X., {et~al.}
  2019{\natexlab{a}}, \mnras, 482, 3162

\bibitem[{{Arrigoni Battaia} {et~al.}(2018){Arrigoni Battaia}, {Prochaska},
  {Hennawi}, {Obreja}, {Buck}, {Cantalupo}, {Dutton}, \&
  {Macci{\`o}}}]{fabrizio18}
{Arrigoni Battaia}, F., {Prochaska}, J.~X., {Hennawi}, J.~F., {et~al.} 2018,
  \mnras, 473, 3907

\bibitem[{{Arrigoni Battaia} {et~al.}(2019{\natexlab{b}}){Arrigoni Battaia},
  {Obreja}, {Prochaska}, {Hennawi}, {Rahmani}, {Ba{\~n}ados}, {Farina}, {Cai},
  \& {Man}}]{fabrizio19b}
{Arrigoni Battaia}, F., {Obreja}, A., {Prochaska}, J.~X., {et~al.}
  2019{\natexlab{b}}, \aap, 631, A18

\bibitem[{{Bacon} {et~al.}(2010){Bacon}, {Accardo}, {Adjali}, {Anwand},
  {Bauer}, {Biswas}, {Blaizot}, {Boudon}, {Brau-Nogue}, \&
  {Brinchmann}}]{bacon10}
{Bacon}, R., {Accardo}, M., {Adjali}, L., {et~al.} 2010, in Society of
  Photo-Optical Instrumentation Engineers (SPIE) Conference Series, Vol. 7735,
  \procspie, 773508

\bibitem[{{Bertone} \& {Schaye}(2012)}]{bertone12}
{Bertone}, S., \& {Schaye}, J. 2012, \mnras, 419, 780

\bibitem[{{Bertone} {et~al.}(2010){Bertone}, {Schaye}, {Booth}, {Dalla
  Vecchia}, {Theuns}, \& {Wiersma}}]{serena10b}
{Bertone}, S., {Schaye}, J., {Booth}, C.~M., {et~al.} 2010, \mnras, 408, 1120

\bibitem[{{Bischetti} {et~al.}(2019){Bischetti}, {Piconcelli}, {Feruglio},
  {Fiore}, {Carniani}, {Brusa}, {Cicone}, {Vignali}, {Bongiorno}, {Cresci},
  {Mainieri}, {Maiolino}, {Marconi}, {Nardini}, \& {Zappacosta}}]{bischetti19}
{Bischetti}, M., {Piconcelli}, E., {Feruglio}, C., {et~al.} 2019, \aap, 628,
  A118

\bibitem[{{Borisova} {et~al.}(2016){Borisova}, {Cantalupo}, {Lilly}, {Marino},
  {Gallego}, {Bacon}, {Blaizot}, {Bouch{\'e}}, {Brinchmann}, {Carollo},
  {Caruana}, {Finley}, {Herenz}, {Richard}, {Schaye}, {Straka}, {Turner},
  {Urrutia}, {Verhamme}, \& {Wisotzki}}]{borisova16}
{Borisova}, E., {Cantalupo}, S., {Lilly}, S.~J., {et~al.} 2016, \apj, 831, 39

\bibitem[{{Bowen} {et~al.}(2016){Bowen}, {Chelouche}, {Jenkins}, {Tripp},
  {Pettini}, {York}, \& {Frye}}]{bowen16}
{Bowen}, D.~V., {Chelouche}, D., {Jenkins}, E.~B., {et~al.} 2016, \apj, 826, 50

\bibitem[{{Cai} {et~al.}(2017){Cai}, {Fan}, {Yang}, {Bian}, {Prochaska},
  {Zabludoff}, {McGreer}, {Zheng}, {Green}, {Cantalupo}, {Frye}, {Hamden},
  {Jiang}, {Kashikawa}, \& {Wang}}]{cai17}
{Cai}, Z., {Fan}, X., {Yang}, Y., {et~al.} 2017, \apj, 837, 71

\bibitem[{{Cai} {et~al.}(2018){Cai}, {Hamden}, {Matuszewski}, {Prochaska},
  {Li}, {Cantalupo}, {Arrigoni Battaia}, {Martin}, {Neill}, {O'Sullivan},
  {Wang}, {Moore}, \& {Morrissey}}]{cai18}
{Cai}, Z., {Hamden}, E., {Matuszewski}, M., {et~al.} 2018, \apjl, 861, L3

\bibitem[{{Cano-D{\'\i}az} {et~al.}(2012){Cano-D{\'\i}az}, {Maiolino},
  {Marconi}, {Netzer}, {Shemmer}, \& {Cresci}}]{cano-diaz12}
{Cano-D{\'\i}az}, M., {Maiolino}, R., {Marconi}, A., {et~al.} 2012, \aap, 537,
  L8

\bibitem[{{Cantalupo} {et~al.}(2014){Cantalupo}, {Arrigoni-Battaia},
  {Prochaska}, {Hennawi}, \& {Madau}}]{cantalupo14}
{Cantalupo}, S., {Arrigoni-Battaia}, F., {Prochaska}, J.~X., {Hennawi}, J.~F.,
  \& {Madau}, P. 2014, \nat, 506, 63

\bibitem[{{Carniani} {et~al.}(2015){Carniani}, {Marconi}, {Maiolino},
  {Balmaverde}, {Brusa}, {Cano-D{\'\i}az}, {Cicone}, {Comastri}, {Cresci},
  {Fiore}, {Feruglio}, {La Franca}, {Mainieri}, {Mannucci}, {Nagao}, {Netzer},
  {Piconcelli}, {Risaliti}, {Schneider}, \& {Shemmer}}]{carniani15}
{Carniani}, S., {Marconi}, A., {Maiolino}, R., {et~al.} 2015, \aap, 580, A102

\bibitem[{{Carniani} {et~al.}(2017){Carniani}, {Marconi}, {Maiolino},
  {Feruglio}, {Brusa}, {Cresci}, {Cano-D{\'\i}az}, {Cicone}, {Balmaverde},
  {Fiore}, {Ferrara}, {Gallerani}, {La Franca}, {Mainieri}, {Mannucci},
  {Netzer}, {Piconcelli}, {Sani}, {Schneider}, {Shemmer}, \&
  {Testi}}]{carniani17}
---. 2017, \aap, 605, A105

\bibitem[{{Chen} {et~al.}(2019){Chen}, {Johnson}, {Straka}, {Zahedy}, {Schaye},
  {Muzahid}, {Bouch{\'e}}, {Cantalupo}, {Marino}, \& {Wendt}}]{chen19}
{Chen}, H.-W., {Johnson}, S.~D., {Straka}, L.~A., {et~al.} 2019, \mnras, 484,
  431

\bibitem[{{Chisholm} {et~al.}(2018){Chisholm}, {Tremonti}, \&
  {Leitherer}}]{chisholm18}
{Chisholm}, J., {Tremonti}, C., \& {Leitherer}, C. 2018, \mnras, 481, 1690

\bibitem[{{Cicone} {et~al.}(2014){Cicone}, {Maiolino}, {Sturm},
  {Graci{\'a}-Carpio}, {Feruglio}, {Neri}, {Aalto}, {Davies}, {Fiore},
  {Fischer}, {Garc{\'\i}a-Burillo}, {Gonz{\'a}lez-Alfonso}, {Hailey-Dunsheath},
  {Piconcelli}, \& {Veilleux}}]{cicone14}
{Cicone}, C., {Maiolino}, R., {Sturm}, E., {et~al.} 2014, \aap, 562, A21

\bibitem[{{Cicone} {et~al.}(2015){Cicone}, {Maiolino}, {Gallerani}, {Neri},
  {Ferrara}, {Sturm}, {Fiore}, {Piconcelli}, \& {Feruglio}}]{cicone15}
{Cicone}, C., {Maiolino}, R., {Gallerani}, S., {et~al.} 2015, \aap, 574, A14

\bibitem[{{Crain} {et~al.}(2013){Crain}, {McCarthy}, {Schaye}, {Theuns}, \&
  {Frenk}}]{crain13}
{Crain}, R.~A., {McCarthy}, I.~G., {Schaye}, J., {Theuns}, T., \& {Frenk},
  C.~S. 2013, \mnras, 432, 3005

\bibitem[{{Dors} {et~al.}(2017){Dors}, {Arellano-C{\'o}rdova}, {Cardaci}, \&
  {H{\"a}gele}}]{dors17}
{Dors}, O.~L., J., {Arellano-C{\'o}rdova}, K.~Z., {Cardaci}, M.~V., \&
  {H{\"a}gele}, G.~F. 2017, \mnras, 468, L113

\bibitem[{{Dors} {et~al.}(2018){Dors}, {Agarwal}, {H{\"a}gele}, {Cardaci},
  {Rydberg}, {Riffel}, {Oliveira}, \& {Krabbe}}]{dors18}
{Dors}, O.~L., {Agarwal}, B., {H{\"a}gele}, G.~F., {et~al.} 2018, \mnras, 479,
  2294

\bibitem[{{Dors} {et~al.}(2014){Dors}, {Cardaci}, {H{\"a}gele}, \&
  {Krabbe}}]{dors14}
{Dors}, O.~L., {Cardaci}, M.~V., {H{\"a}gele}, G.~F., \& {Krabbe}, {\^A}.~C.
  2014, \mnras, 443, 1291

\bibitem[{{Dors} {et~al.}(2019){Dors}, {Monteiro}, {Cardaci}, {H{\"a}gele}, \&
  {Krabbe}}]{dors19}
{Dors}, O.~L., {Monteiro}, A.~F., {Cardaci}, M.~V., {H{\"a}gele}, G.~F., \&
  {Krabbe}, A.~C. 2019, \mnras, 486, 5853

\bibitem[{{Drake} {et~al.}(2019){Drake}, {Farina}, {Neeleman}, {Walter},
  {Venemans}, {Banados}, {Mazzucchelli}, \& {Decarli}}]{drake19}
{Drake}, A.~B., {Farina}, E.~P., {Neeleman}, M., {et~al.} 2019, arXiv e-prints,
  arXiv:1906.07197

\bibitem[{{Farina} {et~al.}(2017){Farina}, {Venemans}, {Decarli}, {Hennawi},
  {Walter}, {Ba{\~n}ados}, {Mazzucchelli}, {Cantalupo}, {Arrigoni-Battaia}, \&
  {McGreer}}]{farina17}
{Farina}, E.~P., {Venemans}, B.~P., {Decarli}, R., {et~al.} 2017, \apj, 848, 78

\bibitem[{{Farina} {et~al.}(2019){Farina}, {Arrigoni-Battaia}, {Costa},
  {Walter}, {Hennawi}, {Drake}, {Decarli}, {Gutcke}, {Mazzucchelli},
  {Neeleman}, {Georgiev}, {Eilers}, {Davies}, {Ba{\~n}ados}, {Fan}, {Onoue},
  {Schindler}, {Venemans}, {Wang}, {Yang}, {Rabien}, \& {Busoni}}]{farina19}
{Farina}, E.~P., {Arrigoni-Battaia}, F., {Costa}, T., {et~al.} 2019, \apj, 887,
  196

\bibitem[{{Fluetsch} {et~al.}(2019){Fluetsch}, {Maiolino}, {Carniani},
  {Marconi}, {Cicone}, {Bourne}, {Costa}, {Fabian}, {Ishibashi}, \&
  {Venturi}}]{fluetsch19}
{Fluetsch}, A., {Maiolino}, R., {Carniani}, S., {et~al.} 2019, \mnras, 483,
  4586

\bibitem[{{Genel} {et~al.}(2014){Genel}, {Vogelsberger}, {Springel}, {Sijacki},
  {Nelson}, {Snyder}, {Rodriguez-Gomez}, {Torrey}, \& {Hernquist}}]{genel14}
{Genel}, S., {Vogelsberger}, M., {Springel}, V., {et~al.} 2014, \mnras, 445,
  175

\bibitem[{{Gilli} {et~al.}(2000){Gilli}, {Maiolino}, {Marconi}, {Risaliti},
  {Dadina}, {Weaver}, \& {Colbert}}]{gilli00}
{Gilli}, R., {Maiolino}, R., {Marconi}, A., {et~al.} 2000, \aap, 355, 485

\bibitem[{{Ginolfi} {et~al.}(2018){Ginolfi}, {Maiolino}, {Carniani}, {Arrigoni
  Battaia}, {Cantalupo}, \& {Schneider}}]{ginolfi18}
{Ginolfi}, M., {Maiolino}, R., {Carniani}, S., {et~al.} 2018, \mnras, 476, 2421

\bibitem[{{Gutkin} {et~al.}(2016){Gutkin}, {Charlot}, \& {Bruzual}}]{gutkin16}
{Gutkin}, J., {Charlot}, S., \& {Bruzual}, G. 2016, \mnras, 462, 1757

\bibitem[{{Hafen} {et~al.}(2019){Hafen}, {Faucher-Gigu{\`e}re},
  {Angl{\'e}s-Alc{\'a}zar}, {Stern}, {Kere{\v{s}}}, {Hummels}, {Esmerian},
  {Garrison-Kimmel}, {El-Badry}, {Wetzel}, {Chan}, {Hopkins}, \&
  {Murray}}]{hafen19}
{Hafen}, Z., {Faucher-Gigu{\`e}re}, C.-A., {Angl{\'e}s-Alc{\'a}zar}, D.,
  {et~al.} 2019, \mnras, 488, 1248

\bibitem[{{Henden} {et~al.}(2018){Henden}, {Puchwein}, {Shen}, \&
  {Sijacki}}]{henden18}
{Henden}, N.~A., {Puchwein}, E., {Shen}, S., \& {Sijacki}, D. 2018, \mnras,
  479, 5385

\bibitem[{{Henden} {et~al.}(2019{\natexlab{a}}){Henden}, {Puchwein}, \&
  {Sijacki}}]{henden19}
{Henden}, N.~A., {Puchwein}, E., \& {Sijacki}, D. 2019{\natexlab{a}}, \mnras,
  2230

\bibitem[{{Henden} {et~al.}(2019{\natexlab{b}}){Henden}, {Puchwein}, \&
  {Sijacki}}]{henden19b}
---. 2019{\natexlab{b}}, arXiv e-prints, arXiv:1911.12367

\bibitem[{{Hennawi} \& {Prochaska}(2013)}]{joseph13}
{Hennawi}, J.~F., \& {Prochaska}, J.~X. 2013, \apj, 766, 58

\bibitem[{{Humphrey} {et~al.}(2011){Humphrey}, {Buote}, {Canizares}, {Fabian},
  \& {Miller}}]{humphrey11}
{Humphrey}, P.~J., {Buote}, D.~A., {Canizares}, C.~R., {Fabian}, A.~C., \&
  {Miller}, J.~M. 2011, \apj, 729, 53

\bibitem[{{Lau} {et~al.}(2016){Lau}, {Prochaska}, \& {Hennawi}}]{lau16}
{Lau}, M.~W., {Prochaska}, J.~X., \& {Hennawi}, J.~F. 2016, \apjs, 226, 25

\bibitem[{{Leclercq} {et~al.}(2017){Leclercq}, {Bacon}, {Wisotzki}, {Mitchell},
  {Garel}, {Verhamme}, {Blaizot}, {Hashimoto}, {Herenz}, {Conseil},
  {Cantalupo}, {Inami}, {Contini}, {Richard}, {Maseda}, {Schaye}, {Marino},
  {Akhlaghi}, {Brinchmann}, \& {Carollo}}]{lseclercq17}
{Leclercq}, F., {Bacon}, R., {Wisotzki}, L., {et~al.} 2017, \aap, 608, A8

\bibitem[{{Lehner} {et~al.}(2015){Lehner}, {Howk}, \& {Wakker}}]{lehner15}
{Lehner}, N., {Howk}, J.~C., \& {Wakker}, B.~P. 2015, \apj, 804, 79

\bibitem[{{Lusso} {et~al.}(2019){Lusso}, {Fumagalli}, {Fossati}, {Mackenzie},
  {Bielby}, {Arrigoni Battaia}, {Cantalupo}, {Cooke}, {Cristiani}, {Dayal},
  {D'Odorico}, {Haardt}, {Lofthouse}, {Morris}, {Peroux}, {Prichard},
  {Rafelski}, {Simcoe}, {Swinbank}, \& {Theuns}}]{lusso19}
{Lusso}, E., {Fumagalli}, M., {Fossati}, M., {et~al.} 2019, \mnras, 485, L62

\bibitem[{{Maiolino} \& {Mannucci}(2019)}]{maiolino19}
{Maiolino}, R., \& {Mannucci}, F. 2019, \aapr, 27, 3

\bibitem[{{Maiolino} {et~al.}(2008){Maiolino}, {Nagao}, {Grazian}, {Cocchia},
  {Marconi}, {Mannucci}, {Cimatti}, {Pipino}, {Fontana}, {Granato},
  {Matteucci}, {Pentericci}, {Risaliti}, {Salvati}, \& {Silva}}]{maiolino08}
{Maiolino}, R., {Nagao}, T., {Grazian}, A., {et~al.} 2008, in Astronomical
  Society of the Pacific Conference Series, Vol. 396, Formation and Evolution
  of Galaxy Disks, ed. J.~G. {Funes} \& E.~M. {Corsini}, 409

\bibitem[{{Maiolino} {et~al.}(2012){Maiolino}, {Gallerani}, {Neri}, {Cicone},
  {Ferrara}, {Genzel}, {Lutz}, {Sturm}, {Tacconi}, {Walter}, {Feruglio},
  {Fiore}, \& {Piconcelli}}]{maiolino12}
{Maiolino}, R., {Gallerani}, S., {Neri}, R., {et~al.} 2012, \mnras, 425, L66

\bibitem[{{Marino} {et~al.}(2019){Marino}, {Cantalupo}, {Pezzulli}, {Lilly},
  {Gallego}, {Mackenzie}, {Matthee}, {Brinchmann}, {Bouch{\'e}}, {Feltre},
  {Muzahid}, {Schroetter}, {Johnson}, \& {Nanayakkara}}]{marino19}
{Marino}, R.~A., {Cantalupo}, S., {Pezzulli}, G., {et~al.} 2019, \apj, 880, 47

\bibitem[{{Martin} {et~al.}(2019){Martin}, {Ho}, {Kacprzak}, \&
  {Churchill}}]{martin19}
{Martin}, C.~L., {Ho}, S.~H., {Kacprzak}, G.~G., \& {Churchill}, C.~W. 2019,
  \apj, 878, 84

\bibitem[{{Maseda} {et~al.}(2017){Maseda}, {Brinchmann}, {Franx}, {Bacon},
  {Bouwens}, {Schmidt}, {Boogaard}, {Contini}, {Feltre}, {Inami},
  {Kollatschny}, {Marino}, {Richard}, {Verhamme}, \& {Wisotzki}}]{maseda17}
{Maseda}, M.~V., {Brinchmann}, J., {Franx}, M., {et~al.} 2017, \aap, 608, A4

\bibitem[{{Matejek} \& {Simcoe}(2012)}]{matejek12}
{Matejek}, M.~S., \& {Simcoe}, R.~A. 2012, \apj, 761, 112

\bibitem[{{Matsuda} {et~al.}(2011){Matsuda}, {Yamada}, {Hayashino}, {Yamauchi},
  {Nakamura}, {Morimoto}, {Ouchi}, {Ono}, {Kousai}, {Nakamura}, {Horie},
  {Fujii}, {Umemura}, \& {Mori}}]{matsuda11}
{Matsuda}, Y., {Yamada}, T., {Hayashino}, T., {et~al.} 2011, \mnras, 410, L13

\bibitem[{{Matsuoka} {et~al.}(2009){Matsuoka}, {Nagao}, {Maiolino}, {Marconi},
  \& {Taniguchi}}]{matsuoka09}
{Matsuoka}, K., {Nagao}, T., {Maiolino}, R., {Marconi}, A., \& {Taniguchi}, Y.
  2009, \aap, 503, 721

\bibitem[{{Matsuoka} {et~al.}(2018){Matsuoka}, {Nagao}, {Marconi}, {Maiolino},
  {Mannucci}, {Cresci}, {Terao}, \& {Ikeda}}]{matsuoka18}
{Matsuoka}, K., {Nagao}, T., {Marconi}, A., {et~al.} 2018, \aap, 616, L4

\bibitem[{{Matsuoka} {et~al.}(2011){Matsuoka}, {Nagao}, {Marconi}, {Maiolino},
  \& {Taniguchi}}]{matsuoka11}
{Matsuoka}, K., {Nagao}, T., {Marconi}, A., {Maiolino}, R., \& {Taniguchi}, Y.
  2011, \aap, 527, A100

\bibitem[{{McIntosh} {et~al.}(1999){McIntosh}, {Rieke}, {Rix}, {Foltz}, \&
  {Weymann}}]{mcintosh99}
{McIntosh}, D.~H., {Rieke}, M.~J., {Rix}, H.~W., {Foltz}, C.~B., \& {Weymann},
  R.~J. 1999, \apj, 514, 40

\bibitem[{{Mignoli} {et~al.}(2019){Mignoli}, {Feltre}, {Bongiorno}, {Calura},
  {Gilli}, {Vignali}, {Zamorani}, {Lilly}, {Le F{\`e}vre}, {Bardelli},
  {Bolzonella}, {Bordoloi}, {Le Brun}, {Caputi}, {Cimatti}, {Diener},
  {Garilli}, {Koekemoer}, {Maier}, {Mainieri}, {Peng}, {P{\'e}rez Montero},
  {Silverman}, \& {Zucca}}]{mignoli19}
{Mignoli}, M., {Feltre}, A., {Bongiorno}, A., {et~al.} 2019, \aap, 626, A9

\bibitem[{{Mingozzi} {et~al.}(2019){Mingozzi}, {Cresci}, {Venturi}, {Marconi},
  {Mannucci}, {Perna}, {Belfiore}, {Carniani}, {Balmaverde}, {Brusa}, {Cicone},
  {Feruglio}, {Gallazzi}, {Mainieri}, {Maiolino}, {Nagao}, {Nardini}, {Sani},
  {Tozzi}, \& {Zibetti}}]{mingozzi19}
{Mingozzi}, M., {Cresci}, G., {Venturi}, G., {et~al.} 2019, \aap, 622, A146

\bibitem[{{Muratov} {et~al.}(2017){Muratov}, {Kere{\v{s}}},
  {Faucher-Gigu{\`e}re}, {Hopkins}, {Ma}, {Angl{\'e}s-Alc{\'a}zar}, {Chan},
  {Torrey}, {Hafen}, {Quataert}, \& {Murray}}]{muratov17}
{Muratov}, A.~L., {Kere{\v{s}}}, D., {Faucher-Gigu{\`e}re}, C.-A., {et~al.}
  2017, \mnras, 468, 4170

\bibitem[{{Nagao} {et~al.}(2006){Nagao}, {Maiolino}, \& {Marconi}}]{nagao06}
{Nagao}, T., {Maiolino}, R., \& {Marconi}, A. 2006, \aap, 447, 863

\bibitem[{{Nakajima} {et~al.}(2018){Nakajima}, {Schaerer}, {Le F{\`e}vre},
  {Amor{\'\i}n}, {Talia}, {Lemaux}, {Tasca}, {Vanzella}, {Zamorani},
  {Bardelli}, {Grazian}, {Guaita}, {Hathi}, {Pentericci}, \&
  {Zucca}}]{nakajima18}
{Nakajima}, K., {Schaerer}, D., {Le F{\`e}vre}, O., {et~al.} 2018, \aap, 612,
  A94

\bibitem[{{Nelson} {et~al.}(2018){Nelson}, {Kauffmann}, {Pillepich}, {Genel},
  {Springel}, {Pakmor}, {Hernquist}, {Weinberger}, {Torrey}, {Vogelsberger}, \&
  {Marinacci}}]{nelson18}
{Nelson}, D., {Kauffmann}, G., {Pillepich}, A., {et~al.} 2018, \mnras, 477, 450

\bibitem[{{Onodera} {et~al.}(2016){Onodera}, {Carollo}, {Lilly}, {Renzini},
  {Arimoto}, {Capak}, {Daddi}, {Scoville}, {Tacchella}, {Tatehora}, \&
  {Zamorani}}]{onodera16}
{Onodera}, M., {Carollo}, C.~M., {Lilly}, S., {et~al.} 2016, \apj, 822, 42

\bibitem[{{Peeples} {et~al.}(2014){Peeples}, {Werk}, {Tumlinson},
  {Oppenheimer}, {Prochaska}, {Katz}, \& {Weinberg}}]{peeples14}
{Peeples}, M.~S., {Werk}, J.~K., {Tumlinson}, J., {et~al.} 2014, \apj, 786, 54

\bibitem[{{P{\'e}rez-Montero} \& {Amor{\'\i}n}(2017)}]{montero17}
{P{\'e}rez-Montero}, E., \& {Amor{\'\i}n}, R. 2017, \mnras, 467, 1287

\bibitem[{{Putman} {et~al.}(2012){Putman}, {Peek}, \& {Joung}}]{putman12}
{Putman}, M.~E., {Peek}, J.~E.~G., \& {Joung}, M.~R. 2012, \araa, 50, 491

\bibitem[{{Rauch} \& {Haehnelt}(2011)}]{rauch11}
{Rauch}, M., \& {Haehnelt}, M.~G. 2011, \mnras, 412, L55

\bibitem[{{Revalski} {et~al.}(2018){Revalski}, {Crenshaw}, {Kraemer},
  {Fischer}, {Schmitt}, \& {Machuca}}]{revalski18}
{Revalski}, M., {Crenshaw}, D.~M., {Kraemer}, S.~B., {et~al.} 2018, \apj, 856,
  46

\bibitem[{{Rubin} {et~al.}(2015){Rubin}, {Hennawi}, {Prochaska}, {Simcoe},
  {Myers}, \& {Lau}}]{rubin15}
{Rubin}, K. H.~R., {Hennawi}, J.~F., {Prochaska}, J.~X., {et~al.} 2015, \apj,
  808, 38

\bibitem[{{Schirmer} {et~al.}(2013){Schirmer}, {Diaz}, {Holhjem}, {Levenson},
  \& {Winge}}]{schirmer13}
{Schirmer}, M., {Diaz}, R., {Holhjem}, K., {Levenson}, N.~A., \& {Winge}, C.
  2013, \apj, 763, 60

\bibitem[{{Shen} {et~al.}(2016){Shen}, {Brandt}, {Richards}, {Denney},
  {Greene}, {Grier}, {Ho}, {Peterson}, {Petitjean}, {Schneider}, {Tao}, \&
  {Trump}}]{shen16}
{Shen}, Y., {Brandt}, W.~N., {Richards}, G.~T., {et~al.} 2016, \apj, 831, 7

\bibitem[{{Sijacki} {et~al.}(2015){Sijacki}, {Vogelsberger}, {Genel},
  {Springel}, {Torrey}, {Snyder}, {Nelson}, \& {Hernquist}}]{sijacki15}
{Sijacki}, D., {Vogelsberger}, M., {Genel}, S., {et~al.} 2015, \mnras, 452, 575

\bibitem[{{Springel}(2010)}]{springel10}
{Springel}, V. 2010, \mnras, 401, 791

\bibitem[{{Trakhtenbrot} \& {Netzer}(2012)}]{trakhtenbrot12}
{Trakhtenbrot}, B., \& {Netzer}, H. 2012, \mnras, 427, 3081

\bibitem[{{Tremonti} {et~al.}(2004){Tremonti}, {Heckman}, {Kauffmann},
  {Brinchmann}, {Charlot}, {White}, {Seibert}, {Peng}, {Schlegel}, {Uomoto},
  {Fukugita}, \& {Brinkmann}}]{tremonti04}
{Tremonti}, C.~A., {Heckman}, T.~M., {Kauffmann}, G., {et~al.} 2004, \apj, 613,
  898

\bibitem[{{Tripp} {et~al.}(2011){Tripp}, {Meiring}, {Prochaska}, {Willmer},
  {Howk}, {Werk}, {Jenkins}, {Bowen}, {Lehner}, {Sembach}, {Thom}, \&
  {Tumlinson}}]{tripp11}
{Tripp}, T.~M., {Meiring}, J.~D., {Prochaska}, J.~X., {et~al.} 2011, Science,
  334, 952

\bibitem[{{Troncoso} {et~al.}(2014){Troncoso}, {Maiolino}, {Sommariva},
  {Cresci}, {Mannucci}, {Marconi}, {Meneghetti}, {Grazian}, {Cimatti},
  {Fontana}, {Nagao}, \& {Pentericci}}]{troncoso14}
{Troncoso}, P., {Maiolino}, R., {Sommariva}, V., {et~al.} 2014, \aap, 563, A58

\bibitem[{{Tumlinson} {et~al.}(2017){Tumlinson}, {Peeples}, \&
  {Werk}}]{tumlinson17}
{Tumlinson}, J., {Peeples}, M.~S., \& {Werk}, J.~K. 2017, \araa, 55, 389

\bibitem[{{Turner} {et~al.}(2014){Turner}, {Schaye}, {Steidel}, {Rudie}, \&
  {Strom}}]{turner14}
{Turner}, M.~L., {Schaye}, J., {Steidel}, C.~C., {Rudie}, G.~C., \& {Strom},
  A.~L. 2014, \mnras, 445, 794

\bibitem[{{van de Voort} \& {Schaye}(2013)}]{freeke13}
{van de Voort}, F., \& {Schaye}, J. 2013, \mnras, 430, 2688

\bibitem[{{Vangioni} {et~al.}(2018){Vangioni}, {Dvorkin}, {Olive}, {Dubois},
  {Molaro}, {Petitjean}, {Silk}, \& {Kimm}}]{vangioni18}
{Vangioni}, E., {Dvorkin}, I., {Olive}, K.~A., {et~al.} 2018, \mnras, 477, 56

\bibitem[{{Venemans} {et~al.}(2017){Venemans}, {Walter}, {Decarli},
  {Ferkinhoff}, {Wei{\ss}}, {Findlay}, {McMahon}, {Sutherland}, \&
  {Meijerink}}]{venemans17}
{Venemans}, B.~P., {Walter}, F., {Decarli}, R., {et~al.} 2017, \apj, 845, 154

\bibitem[{{Villar-Mart{\'\i}n} {et~al.}(2007){Villar-Mart{\'\i}n},
  {S{\'a}nchez}, {Humphrey}, {Dijkstra}, {di Serego Alighieri}, {De Breuck}, \&
  {Gonz{\'a}lez Delgado}}]{villar07}
{Villar-Mart{\'\i}n}, M., {S{\'a}nchez}, S.~F., {Humphrey}, A., {et~al.} 2007,
  \mnras, 378, 416

\bibitem[{{Vogelsberger} {et~al.}(2014){Vogelsberger}, {Genel}, {Springel},
  {Torrey}, {Sijacki}, {Xu}, {Snyder}, {Nelson}, \&
  {Hernquist}}]{vogelsberger14}
{Vogelsberger}, M., {Genel}, S., {Springel}, V., {et~al.} 2014, \mnras, 444,
  1518

\bibitem[{{Wisotzki} {et~al.}(2016){Wisotzki}, {Bacon}, {Blaizot},
  {Brinchmann}, {Herenz}, {Schaye}, {Bouch{\'e}}, {Cantalupo}, {Contini},
  {Carollo}, {Caruana}, {Courbot}, {Emsellem}, {Kamann}, {Kerutt}, {Leclercq},
  {Lilly}, {Patr{\'\i}cio}, {Sandin}, {Steinmetz}, {Straka}, {Urrutia},
  {Verhamme}, {Weilbacher}, \& {Wendt}}]{wisotzki16}
{Wisotzki}, L., {Bacon}, R., {Blaizot}, J., {et~al.} 2016, \aap, 587, A98

\bibitem[{{Witstok} {et~al.}(2019){Witstok}, {Puchwein}, {Kulkarni}, {Smit}, \&
  {Haehnelt}}]{witstok19}
{Witstok}, J., {Puchwein}, E., {Kulkarni}, G., {Smit}, R., \& {Haehnelt}, M.~G.
  2019, arXiv e-prints, arXiv:1905.06954

\bibitem[{{Xu} {et~al.}(2018){Xu}, {Bian}, {Shen}, {Zuo}, {Fan}, \&
  {Zhu}}]{xu18}
{Xu}, F., {Bian}, F., {Shen}, Y., {et~al.} 2018, \mnras, 480, 345

\bibitem[{{Yuma} {et~al.}(2017){Yuma}, {Ouchi}, {Drake}, {Fujimoto}, {Kojima},
  \& {Sugahara}}]{yuma17}
{Yuma}, S., {Ouchi}, M., {Drake}, A.~B., {et~al.} 2017, \apj, 841, 93

\bibitem[{{Yuma} {et~al.}(2019){Yuma}, {Ouchi}, {Fujimoto}, {Kojima}, \&
  {Sugahara}}]{yuma19}
{Yuma}, S., {Ouchi}, M., {Fujimoto}, S., {Kojima}, T., \& {Sugahara}, Y. 2019,
  arXiv e-prints, arXiv:1904.11510

\end{thebibliography}

\end{document}